\documentclass{elsarticle}

\usepackage{xcolor}
\usepackage{hyperref}
\usepackage[export]{adjustbox}
\usepackage{amsmath}
\usepackage{amssymb}
\usepackage{array}
\usepackage{arydshln}

\journal{Applied Ocean Research}

\newcommand{\fig}[1]{Fig.~\ref{#1}}
\newcommand{\eq}[1]{Eq.~(\ref{#1})}
\newcommand{\ga}{\gtrsim}
\newcommand{\la}{\lesssim}
\newcommand{\thickfoil}{w}

\begin{document}

\begin{frontmatter}

\title{A two-dimensional analytical model of vertical water entry for 
asymmetric bodies with flow separation}

\author[enstaAddress]{Romain Hasco\"{e}t}
\cortext[mycorrespondingauthor]{Corresponding author}
\ead{romain.hascoet@ensta-bretagne.fr}

\author[enstaAddress]{Nicolas Jacques}
\author[enstaAddress]{Yves-Marie Scolan}
\author[ifremerAddress]{Alan Tassin}

\address[enstaAddress]{ENSTA Bretagne, CNRS UMR 6027, IRDL, 2 rue Fran\c{c}ois Verny, 29806 Brest Cedex 9, France}
\address[ifremerAddress]{IFREMER -- LCSM, ZI Pointe du Diable, 29280 Plouzan\'{e} CS 10070, France}

\begin{abstract}
The vertical water entry of asymmetric two-dimensional bodies with flow separation is considered.
As long as there is no flow separation, linearised Wagner's theory combined with the Modified Logvinovich Model 
has been shown to provide computationally fast and reliable estimates of slamming loads during water entry.
Tassin et al. (2014) introduced the Fictitious Body Continuation (FBC) concept as a way to extend the use of Wagner's model
to separated flow configurations, but they only considered 
symmetric bodies. 
In the present study, we investigate the ability of the FBC concept to provide accurate estimates of slamming loads
for asymmetric bodies. 
In this case, flow separation may not occur simultaneously on both sides of the body.
During an intermediate phase,
slamming loads are governed by a competition between the local drop in pressure 
due to partial flow separation
and the ongoing expansion of the wetted area.
As a first benchmark for the model, we consider the water entry of an inclined flat plate 
and compare the FBC estimates with the results of a nonlinear model.
Then, we consider the case of a foil and compare the FBC results with Computational Fluid Dynamics predictions.
In both cases, we find that the FBC model is able to provide reliable estimates of the slamming loads. 
\end{abstract}


\end{frontmatter}

\section{Introduction}

Water impacts are complex phenomena which have been extensively studied by means of experiments, 
analytical modelling
and numerical simulations.
From an engineering standpoint, 
it is crucial to know the hydrodynamic loads to which a body is 
exposed
when entering water.
Among the various applications, one can think of ship slamming, 
sloshing in tanks, aircraft ditching, 
water landing
of seaplanes.
Experimental campaigns demand time and specific facilities, and protocols can face technical difficulties (e.g. pressure measurements).
Modern Computational Fluid Dynamics (CFD) models offer hope for a detailed description of the impact-generated flow,
but they are numerically 
demanding
and the validation of results can be problematic; especially during the early stage of the impact.
Alternatively, analytical developments require simplifying assumptions to make the problem tractable,
but provide computationally fast results that are easier to check and control.

The pioneering work of Wagner (1932) \cite{wagner_1932} is still used as the basic framework of many (semi-)analytical 
models of water impact.
One of the main approximation of Wagner's model is the projection of the wetted body surface to the calm-water reference plane:
the so called flat-disc 
or flat-plate 
approximation. 
Thus, Wagner's approach 
is theoretically
restricted to bodies with small deadrise angles.
The flow is assumed to be potential and the speed of penetration is supposed sufficiently high to neglect gravity.
In slamming problems, viscous effects can usually be considered 
negligible
(e.g. \cite{muzaferija_1998}).
The rapid expansion of the wetted area generates splash water jets 
at the line of contact
which delimits the wetted surface of the body (hereafter the \textit{contact line})
\cite{greenhow_1983, lin_1997}. 
These jets connect to the main flow through a root region 
where the curvature of the free surface is high and the flow is strongly nonlinear. 
Slamming pressure peaks in the root region, and quickly decreases toward the tip of the jet.
Except for the root region, the water jets do not significantly contribute to the slamming loads \cite{cointe_1989}.
Wagner's problem is valid only if the wetted region is expanding \cite{howison_1991}; 
i.e. there can be no migration of fluid particles from the wetted area to the free surface.
Therefore, the classical Wagner model is not suitable to study water exit or the development of separated flows during water entry.

Flow separation may occur when the contact line 
reaches body knuckles, or smooth 
convex regions where the local deadrise angle 
reaches sufficiently large values.
Then, a cavity flow forms behind the body and hydrodynamic loads usually start decreasing.
It may be important to know how fast slamming pressure decays to predict the transient and vibratory response 
of the body structure. 
Besides, for asymmetric bodies or oblique entries, flow separation may not happen all at once over the contact line. 
Then there will be a transient phase where 
the evolution of slamming loads will be governed by a competition between the local drop in pressure due to flow separation
and the ongoing expansion of wetted area.
The water entry of finite wedges (flow separation at knuckles) has been extensively studied 
by means of experiments \cite{greenhow_1987, tveitnes_2008, wang_2015},
analytical developments \cite{logvinovich_1972, tassin_2014, duan_2013}, 
fully nonlinear potential simulations \cite{zhao_1996, iafrati_2003, wang_2015,bao_2016,bao_2017}, Navier-Stokes simulations \cite{maki_2011,piro_2013,gu_2014}
and smoothed-particle hydrodynamics method \cite{oger_2006}.
When flow separation occurs on a smooth part of the body, 
the location of the separation line is not known \textit{a priori}, and it may evolve in time.
The separation location can be sensitive to various parameters such as the impact velocity and the wettability properties of the body surface 
 (see for example \cite{worthington_1908, lin_1997, duez_2007}). 
In hydrodynamic models of water entry, 
the microphysics of surface interactions -- between water, air and the solid -- is ignored. 
Flow separation on a smooth body is governed by pressure interaction between air and water at the contact line (\textit{nonviscous} separation), 
as illustrated by the CFD simulations of Zhu, Faltinsen and Hu (2006) \cite{zhu_2006}.
When the fluid pressure in the jet root region drops below air pressure, air can seep in between the body and the main flow,
leading to separation.
Sun and Faltinsen (2006, 2007) \cite{sun_2006, sun_2007} used this criterion to 
trigger
flow separation in a boundary element method (BEM) scheme and 
obtained good agreement with experiments on the separation location and the free surface of the early separated flow
during the vertical water entry of a circular cylinder.

Inspired by the work of Logvinovich (1972) \cite{logvinovich_1972} and more recent publications \cite{fairlie_2008, malenica_2007, duan_2013},
Tassin et al. (2014) \cite{tassin_2014} investigated the `Fictitious Body Continuation' (FBC) concept 
as an effective way to extend the use of Wagner's model after flow separation from the body.
The principle of the FBC model is to extend the real body by a fictitious one so that Wagner's model can be applied to the composite real+fictitious body.
The pressure along the composite body contour is estimated by using the Modified Logvinovich Model (MLM), introduced by Korobkin (2004) \cite{korobkin_2004}.
Then, the hydrodynamic load is obtained by integrating the pressure along the real part of the body only.
Within this approach, the main question to address is whether there exists 
a simple and generic fictitious body shape that can properly mimic the early expansion of the cavity flow behind the body.
Tassin et al. \cite{tassin_2014} have given the first part of the answer by considering simple symmetric bodies: 
a horizontal flat plate, wedges with different deadrise angles and a circular cylinder.
By comparing the FBC estimates with experimental and CFD results, 
they 
found that a continuation with inclined flat plates 
could
give good agreements 
on the hydrodynamic loads during the early stage of cavity 
initiation.
They 
found as `best-fit' continuation angles, $\alpha \simeq 47^{\circ}$ for 
a
flat plate, 
$\alpha \sim 45^{\circ}-55^{\circ}$ for wedges with deadrise angles between $10^{\circ}$ and $30^{\circ}$,
and $\alpha \simeq 60^{\circ}$ for the circular cylinder.
Interestingly, these continuation angles obtained for quite different shapes do not spread over a large range.

In the present paper, we go one step further and investigate whether the continuation by inclined flat plates can still provide satisfactory results 
for \textit{asymmetric} bodies. 
Section \ref{sect_analytical_model} introduces the theoretical framework.
In Section \ref{sec_inclined_fp}, we first consider the case of an inclined flat plate 
and compare FBC estimates with the nonlinear self-similar model of Faltinsen and Semenov (2008) \cite{faltinsen_semenov_2008}.
In Section \ref{sec_naca}, we test the model capabilities on the vertical water entry of a foil, 
which is an asymmetric body mixing flow separations at a chine and on a smooth part of the body contour.
FBC estimates are compared with CFD results obtained from simulations carried out with the finite-element software ABAQUS/Explicit.
As FBC estimates are found to be reliable, we also give a qualitative and quantitative description of slamming loads 
on foils with different thicknesses.
The model is further discussed in Section \ref{sect_conclusions}.

\section{Analytical model}
\label{sect_analytical_model}

\subsection{Wagner's model for a 2D asymmetric body}

\begin{figure}[h]
\begin{center}
\begin{tabular}{c}
        \includegraphics[width=0.9\textwidth]{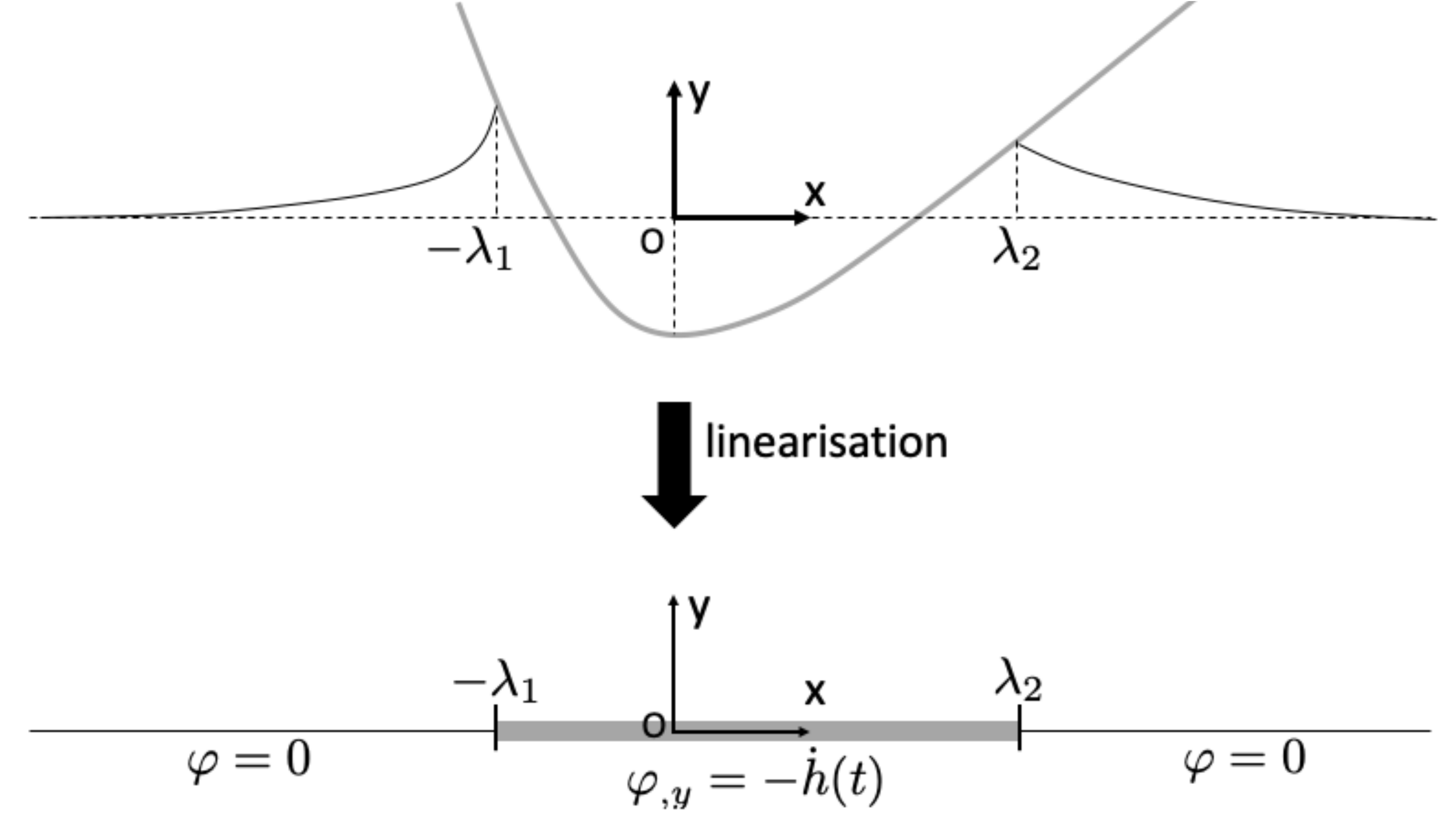}
\end{tabular}
\end{center}
\caption{
Illustration of Wagner's model for vertical water entry.
The boundary conditions are linearised and projected onto the initial free surface plane. 
The projection transforms the flow around the real body shape 
into the flow past a flat plate (both represented in grey).
The size of the contact region is determined via the Wagner condition,
which states that the vertical displacement is 
finite
at $x = -\lambda_1, \lambda_2$.
}
\label{fig_wagner_problem}
\end{figure}

In Wagner's model, the fluid is considered inviscid and incompressible. 
Gravity and surface tension effects are neglected (see appendix A for a discussion about the validity of the no gravity assumption).
The flow is assumed to be irrotational, so the flow velocity $\textbf{V}$ can be expressed by the gradient of a scalar potential $\varphi$.
Before its first contact with the body, the fluid is initially at rest and its domain is delimited by a flat free surface.
The problem is formulated into a fixed 
coordinate system
$Oxy$ 
whose origin coincides with the location of the first fluid-body contact point (see \fig{fig_wagner_problem}).
The body contour can be represented by a continuous shape function $f(x)$ satisfying $f(0) = 0$.
At a given time, $t$, the location of the body contour is given by $y(x,t) = f(x) - h(t)$, 
where $h$ is the vertical penetration depth of the body; $h$ is positive and $h(0) = 0$.
The local deadrise angle\footnote{
In the present paper, $Q_{,v}$ denotes the derivative of the function $Q$ with respect to the variable $v$.
}, $\beta(x) =  {\rm atan} (f_{,x}(x))$, between the body contour and 
the initial free surface is assumed to be small,  
$\beta \simeq f_{,x} \ll 1$.
Following these assumptions, 
the mixed boundary value problem satisfied by the velocity potential can be linearised as follows
(see \fig{fig_wagner_problem} for illustration):

\begin{align}
 \label{eq_laplace} \Delta \varphi  & =    0 \, , & &  y<0  \\
 \label{eq_free_kin} \varphi  & =   0 \, ,  & & y=0, x \in ] -\infty , - \lambda_1] \cup [\lambda_2,  + \infty [ \\
 \label{eq_imp} \varphi_{,y} & =  -\dot{h}(t) \, , & & y=0, x \in ]\lambda_1, \lambda_2[ \\
 \label{eq_far} \varphi & \rightarrow 0 \, , &  & x^2 + y^2 \rightarrow \infty
\end{align}
where $\dot{h}$ is the time derivative of $h$ (i.e. the vertical entry velocity).
\eq{eq_laplace} is Laplace's equation restricted to the \textit{initial} fluid domain.
\eq{eq_free_kin} is the linearised dynamic boundary condition of the free surface projected onto its initial plane.
\eq{eq_imp} is the impermeability condition expressed for the equivalent flat plate.
\eq{eq_far} is the far-field condition.
Although all equations are linearised and projected onto a fixed plane, 
the water entry problem
remains nonlinear as the wetted region $[-\lambda_1 , \lambda_2]$ is unknown \textit{a priori}.
The wetted area has to be determined together with the flow solution 
by imposing the so called Wagner condition at the fluid-body contact points
\begin{align}
\label{eq_wag_cond_1} \eta(-\lambda_1,t) & = f(-\lambda_1) - h(t) \, , \\
\label{eq_wag_cond_2} \eta(\lambda_2,t) & = f(\lambda_2) - h(t) \, ,
\end{align}
where $\eta(x,t)$ is the free surface elevation, 
and $h(t)$ the penetration depth.
Using the linearised kinematic condition, the free-surface elevation is obtained from 
\begin{equation}
\label{eq_kinema}
\eta (x,t) = \int_{0}^{t} \varphi_{,y}(x,y=0,\tau) \ {\rm d} \tau \, ,
\end{equation}
which closes the problem. 
Eq. (\ref{eq_laplace}-\ref{eq_kinema}) form Wagner's problem.

By expressing Wagner's problem 
for the displacement potential
\begin{equation}
\Phi(x,y,t) = \int_0^t \varphi(x,y,\tau) d\tau \, ,
\end{equation}
and enforcing finite displacements at the contact points,
Wagner's condition can be reduced to a system of two nonlinear equations (see \cite{scolan_1999} for details): 
\begin{align}
\displaystyle \int_{-\lambda_1}^{\lambda2} f(x) \sqrt{\frac{\lambda_2-x}{\lambda_1+x}} \ {\rm d} x & = \displaystyle \frac{\pi}{2} (\lambda_1 + \lambda_2) h \label{wagner_condi_2d_left} \, , \\
\displaystyle \int_{-\lambda_1}^{\lambda2} f(x) \sqrt{\frac{\lambda_1+x}{\lambda_2-x}} \ {\rm d} x & = \displaystyle \frac{\pi}{2} (\lambda_1 + \lambda_2) h \label{wagner_condi_2d_right} \, .
\end{align}
The penetration depth $h$ enters Eqs. (\ref{wagner_condi_2d_left}-\ref{wagner_condi_2d_right}) as a parameter, 
which implies that the wetted area only depends on the current position of the body;
the history of body kinematics has no influence.
If $f(x)$ is a polynomial of low degree, the system of equations (\ref{wagner_condi_2d_left}-\ref{wagner_condi_2d_right}) can be analytically solved for $\lambda_1$ and $\lambda_2$
\cite{cointe_2004}.
In the case of arbitrary shapes, 
these equations can be numerically solved by using a root-finding algorithm.
In the present study, we use Newton's method with a relaxation condition
(see appendix B for details).

\subsection{Pressure}

Knowing the expansion of the Wagner wetted area, 
the pressure can be computed by means of different models. 
The simplest one is the linearised Bernoulli equation 
\begin{equation}
\label{eq_p_linear}
P (x,t)  = - \rho \varphi_{,t}^{(w)}(x,0,t) \, ,
\end{equation}
where $\rho$ is the fluid density and $\varphi^{(w)}$ is the velocity potential solution of Wagner's problem. 
For vertical water entry, $\varphi^{(w)}$ is the velocity potential on a flat plate:
\begin{equation}
\label{eq_plaque_plane}
\varphi^{(w)}(x,0,t) = -\dot{h}(t) \sqrt{\left[\lambda_1(t)+x\right]\left[\lambda_2(t)-x\right]} \, .
\end{equation} 
This linear model has been shown to overpredict hydrodynamics loads (see \cite{faltinsen_2006} for a detailed discussion).
In order to improve pressure predictions, the quadratic term of Bernoulli's equation has to be taken into  account. 
However, the quadratic term $[\varphi_{,x}^{(w)}]^2$ is not integrable at the contact points.  
To circumvent this difficulty, the Wagner flow solution can be asymptotically matched with a solution of the jet flow emerging
from the root region \cite{cointe_1987, cointe_1989, howison_1991, zhao_1993}.

As an alternative, in the present work, 
we 
use the Modified Logvinovich Model (MLM), 
proposed
more recently by 
Korobkin (2004) \cite{korobkin_2004}.
The MLM model takes into account the quadratic term and the shape of the body for the pressure calculation, 
by making use of the exact Bernoulli equation expressed at the body contour:
\begin{equation}
\label{eq_pmlm_general}
P(x,t) = - \rho \left[ \phi_{,t} + \frac{f_{,x}\dot{h}}{1+f_{,x}^2} \phi_{,x} + \frac{1}{2(1+f_{,x}^2)} (\phi_{,x}^2 - \dot{h}^2) \right] \, ,
\end{equation}
where $P$ and $\phi$ are the pressure and the velocity potential along the body contour:
\begin{equation}
\begin{array} {ccc}
P(x,t) & = & p(x,f(x)-h(t),t) \\
\phi(x,t) & = & \varphi(x,f(x)-h(t),t) \, .
\end{array}
\end{equation}
Then, 
the velocity potential $\phi$ is approximated by a first-order Taylor expansion of the Wagner solution,
 from the initial free surface level to the body height:
\begin{equation}
\label{eq_vel_pot_mlm}
\phi(x,t) \simeq \varphi^{(w)}(x,0,t) - \dot{h}(t)(f(x)-h(t)) \, .
\end{equation}
Thus part of the second order correction for the velocity potential is taken into account.
Note however that the second-order perturbative component of the velocity potential is ignored in the MLM approach
; see \cite{korobkin_2007,oliver_2007} for a second-order analysis of Wagner's model.
Substituting 
\eq{eq_vel_pot_mlm} into \eq{eq_pmlm_general}, the MLM pressure for the vertical water entry of a 2D asymmetric body reads (see also \cite{korobkin_2005}):
\begin{eqnarray}
\label{eq_pmlm}
P(x,t) & = &  \underbrace{ \frac{1}{2} \rho \dot{h}^2 \left[ \frac{{\rm d}\lambda_2}{{\rm d}h} \sqrt{\frac{\lambda_1+x}{\lambda_2-x}} +  
\frac{{\rm d}\lambda_1}{{\rm d}h} \sqrt{\frac{\lambda_2-x}{\lambda_1+x}} 
- \frac{1}{4} \frac{(\lambda_2-\lambda_1-2x)^2}{(\lambda_2-x)(\lambda_1+x)(1+f_x^2)} - 1\right]}_{P_v(x,t)} \nonumber \\
& & + \underbrace{ \rho \ddot{h} \left[ \sqrt{(\lambda_2-x)(\lambda_1+x)}+f(x)-h(t) \right] }_{P_a(x,t)} \, .
\end{eqnarray}
The first and second terms of \eq{eq_pmlm}, $P_v$ and $P_a$, respectively represent the slamming and added-mass pressures.
%}

As the quadratic MLM pressure term $\phi_x^2$ has non-integrable singularities 
close to the contact points, an additional condition has to be introduced to make the model effective.
From an idea already expressed by Wagner \cite{wagner_1932}, 
Korobkin \cite{korobkin_2004} suggested to ignore regions of negative pressure close to the contact points.
Negative pressure regions need to be ignored for the slamming term only,
as the added-mass term is not singular at contact points.

Although this last 
condition
lacks a physically grounded justification, 
the MLM model has shown good agreement with CFD simulations and experiments regarding the hydrodynamic force and the pressure distribution.
The MLM model has been shown to be accurate up to relatively
``large'' deadrise angles
(see \cite{korobkin_2004, korobkin_2005, tassin_2010}),
beyond the original validity domain of Wagner's model.

\vspace{-0.1cm}
\subsection{Fictitious Body Continuation}
\label{subsec_fbc_cont}

Tassin et al. \cite{tassin_2014} 
used
the Fictitious Body Continuation concept 
to extend the use of Wagner's model after flow separation from the body.
Inspired by an original idea from Logvinovich (1972) \cite{logvinovich_1972}, 
the principle of the FBC model is to extend the real body by a fictitious one so that Wagner's model can be applied to the composite real+fictitious body.
Then the hydrodynamic loads are obtained by integrating the MLM pressure along the real part of the body only.
After flow separation, CFD simulations \cite{seng_2012} indeed suggest that the hydrodynamic drag 
can still be decomposed into a velocity component proportional to $\dot{h}^2$ and an acceleration component proportional to $\ddot{h}$. 
As the added-mass pressure, $P_a$, is expected to level off after flow separation,
Tassin et al. \cite{tassin_2014} modified the added-mass term of \eq{eq_pmlm}, 
as follows
\vspace{-0.1cm}
\begin{equation}
\label{eq_pa_sym}
P_a(x,t) = \rho \ddot{h} \left[ f(x) - {\rm min}[h(t),f(l)] + \sqrt{{\rm min}(\lambda,l)^2-x^2} \right] \, ,
\end{equation}
for a symmetric body, where $\lambda$ is the half-width of the wetted area on the composite body
and $\pm l$ are the abscissa of the separation points.
The presence of the minimum operators ensures that \eq{eq_pa_sym} is valid before and after flow separation.
In the case of asymmetric bodies (see \fig{fig_naca_sf} for an illustration), this 
modification
has to be generalised 
because separation does not occur at the same height and same time on both sides of the body.
We suggest to use
\vspace{-0.1cm}
\begin{equation}
P_a(x,t) = \rho \ddot{h} \left[ f(x) - {\rm min}(h(t),\tilde{f}(x)) + \sqrt{ ({\rm min}(\lambda_1,l_1)+x) ({\rm min}(\lambda_2,l_2)-x) } \right]
\end{equation}
where $-l_1$ and $l_2$ are the abscissa of the separation points on the left and on the right respectively.
When the flow separation occurs on a smooth part of the body, 
the location of the separation point depends on the shape chosen for the fictitious continuation 
(see \S\ref{subsec_naca_geom} for the case study of a foil).
$\tilde{f}$ is a linear interpolation between the two separation heights 
\vspace{-0.1cm}
\begin{equation}
\tilde{f} (x) = f(-l_1) + \frac{x+l_1}{l_2+l_1} \left[f(l_2)-f(-l_1)\right] \, .
\end{equation}
In \S\ref{par_abaqus_acc}, 
we show that this simple prescription provides satisfactory agreement with CFD results.

%%%%%%%%%%%%%%%%%%%%%%%%%%%%%%%%%%%%%%%

\section{
Water entry of
an inclined flat plate}
\label{sec_inclined_fp}

% %%%%%%%%%%%%%%%%%%%%
% %%% FIGURES %%%%%%%%%%%
% %%%%%%%%%%%%%%%%%%%%

\begin{figure}[t]
\begin{center}
\begin{tabular}{c}
        \includegraphics[width=0.9\textwidth]{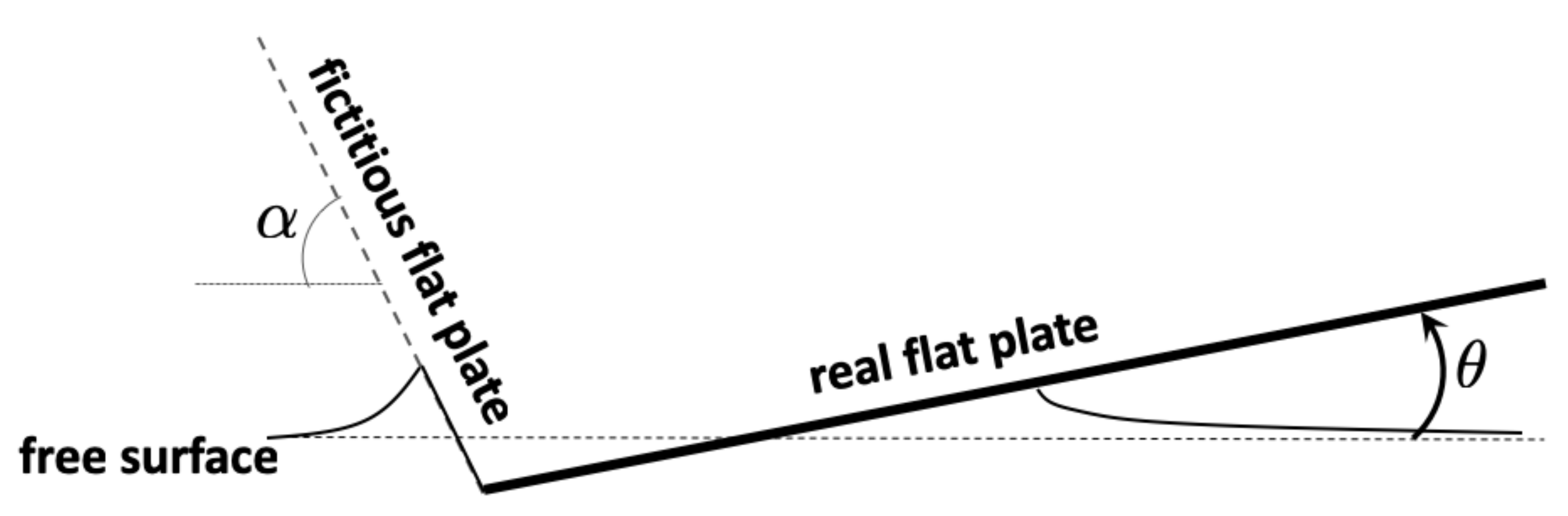}    
\end{tabular}
\end{center}
\caption{
Vertical water entry of a flat plate -- illustration.
The flat plate is inclined by an angle $\theta$.
Flow separation on the left occurs right at the beginning of the water entry.
In the FBC model the separated flow is mimicked by a fictitious flat plate with inclination $\alpha$;
the real and fictitious flat plates form an inclined wedge.
}
\label{fig_flat_plate_sketch}
\end{figure}

\begin{figure}[h]
\begin{center}
\begin{tabular}{c}
        \includegraphics[width=0.5\textwidth]{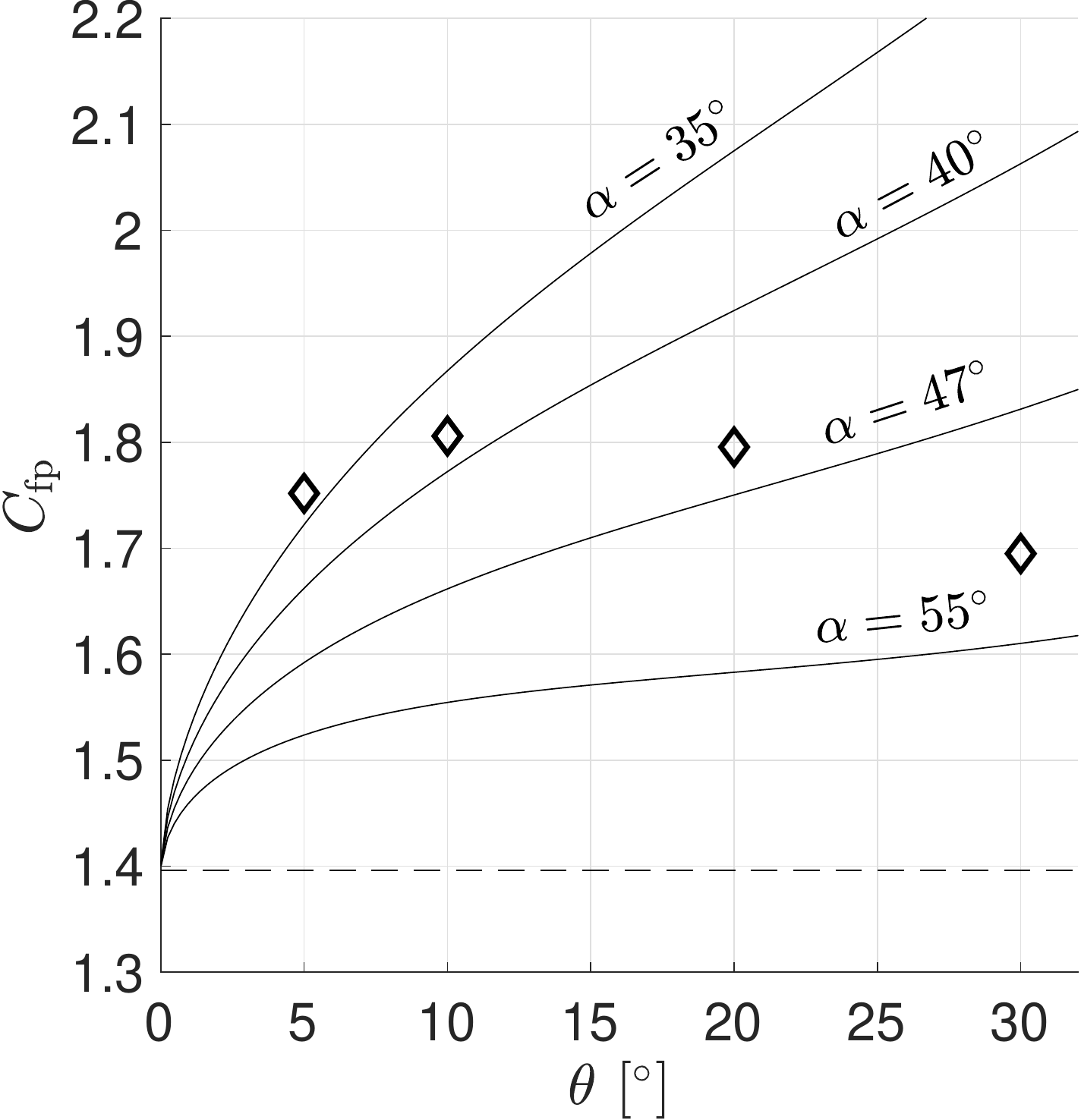}    
\end{tabular}
\end{center}
\caption{
Vertical water entry of an inclined flat plate at constant velocity: 
normalised vertical force $C_{\rm fp}$ (see Eq. \ref{eq_cfp}) as a function of the plate inclination angle $\theta$.
Solid lines show the predictions of the FBC model for different continuation angles $\alpha$.
The dashed line shows the prediction of the classical Wagner model 
with the use of linearised Bernoulli's equation for pressure integration (Eq. \ref{eq_fy_fp}).
Diamonds show results from the nonlinear self-similar model of Faltinsen and Semenov \cite{faltinsen_semenov_2008}.
\vspace{-0.25cm}
}
\label{fig_flat_plate}
\end{figure}

% %%%%%%%%%%%%%%%%%%%%
% %%% FIGURES %%%%%%%%%%%
% %%%%%%%%%%%%%%%%%%%%

\subsection{Classical Wagner model with linearised Bernoulli equation}
\label{subsec_fp_wagner}

The vertical water entry of an inclined flat plate can be investigated with the classical Wagner model
by assuming that the free surface of the separated flow expands vertically\footnote{
In reality the separated flow free surface will expand in a more complicated shape;
see for example \cite{faltinsen_semenov_2008}.}.
Within the Fictitious Body Continuation concept, this corresponds to the situation where the real flat plate is continued 
from its leading edge 
by a vertical fictitious flat plate, with inclination angle $\alpha= 90^{\circ}$ (see \fig{fig_flat_plate_sketch}).
It is a limit case for Wagner's model as the wetted area does not expand on one side of the body, 
leading to a singular free surface at the plate leading edge ($x=0$).
Thus the contact point location on the left is known \textit{a priori}, $\lambda_1 = 0$, 
and the corresponding Wagner condition (Eq. \ref{wagner_condi_2d_left}) does not hold.
The location of the other contact point is given by \eq{wagner_condi_2d_right}, setting $\lambda_1 = 0$:
\begin{equation}
\label{eq_wetted_corr_fp}
{\lambda_2}^{(w)} = \frac{4}{3}\frac{h}{\tan \theta} \, .
\end{equation}
Reinhard \cite{reinhard_2013} obtained the same expression for ${\lambda_2}^{(w)}$
by using a slightly different approach: instead of imposing finite displacement at the contact point, 
he required the displacement potential to satisfy the far-field condition.
For the sake of comparison
(see the following paragraph), 
it is then interesting to give the resulting vertical force component obtained by integrating the linearised Bernoulli relation 
(Eq. \ref{eq_p_linear}):
\begin{align}
F_y & =  \frac{\pi}{4} \rho \dot{h} \dot{\lambda}_2 \lambda_2  + \frac{\pi}{8} \rho \ddot{h} \lambda_2^2 \nonumber \\
       & =  \frac{4\pi}{9} \frac{h}{\tan^2\theta} \rho \dot{h}^2 + \frac{2\pi}{9} \frac{h^2}{\tan^2\theta} \rho \ddot{h} \label{eq_fy_fp} \, .
\end{align}

\subsection{Fictitious Body Continuation compared with a nonlinear model}

The vertical force component acting on an inclined flat plate entering water
can be normalised to a coefficient
\vspace{-0.1cm}
\begin{equation}
\label{eq_cfp}
C_{\rm fp} = \frac{\tan^2\theta}{h} \frac{F_y }{\rho \dot{h}^2} \, ,
\end{equation}
where the factor $\tan^2\theta$ has been added to limit the range of values.
Flow separation occurs at the leading edge, right at the first fluid-solid contact.
This configuration can be addressed with the FBC model by connecting a fictitious flat plate to the leading edge of the real flat plate (see \fig{fig_flat_plate_sketch}). 
Within this framework, the flow induced by the inclined flat plate is self-similar 
and $C_{\rm fp}$ does not depend on the penetration depth $h$ if the entry velocity is constant.

\fig{fig_flat_plate} shows the force coefficient $C_{\rm fp}$ predicted by the FBC model, 
as a function of the plate inclination angle $\theta$, for constant entry velocity.
Estimates are plotted for different continuation angles of the fictitious flat plate.
To assess the relevancy of the FBC concept, our results are compared with 
calculations from the nonlinear self-similar model of Faltinsen and Semenov (see Table 2 in \cite{faltinsen_semenov_2008}).
For a continuation angle $\alpha=47^{\circ}$, both models agree by $\sim 10\%$ over a range of inclination angles $\theta=5^{\circ}\rightarrow30^{\circ}$.
This is consistent with the results obtained by Tassin et al. (2014) \cite{tassin_2014}
for the vertical water entry of a horizontal flat plate:
by comparing with the numerical calculations of Iafrati and Korobkin (2008) \cite{iafrati_2008}, 
they found that the FBC model with $\alpha=47^{\circ}$ 
can reproduce the early decay of the hydrodynamic force acting on the plate.
The force coefficient  $C_{\rm fp}$ predicted by the nonlinear model of Faltinsen and Semenov starts decreasing for $\theta \ga 20^{\rm \circ}$,
while the values obtained from the FBC model with $\alpha=47^{\circ}$ keep increasing.
In principle, it should be possible to find a ``phenomenological'' law $\alpha(\theta)$ 
for which agreement between both models remains 
good for $\theta \ga 30^{\circ}$.
We do not attempt to give such a law in the present paper.

For $\theta \rightarrow 0$, all FBC curves converge to the same coefficient value $C_{\rm fp} \simeq 1.4$. 
This asymptotic value coincides with the force coefficient predicted by the classical Wagner model,
$C_{\rm fp} = 4\pi/9$ (see Eq. \ref{eq_fy_fp}).
This can be understood as follows. 
First as $\theta \rightarrow 0$, the wetted area on the real plate expands much faster than on the fictitious one, with $\lambda_1 / \lambda_2 \rightarrow 0$,
resulting in $\lambda_2 \rightarrow {\lambda_2}^{(w)}$.
Then, an asymptotic analysis of \eq{eq_pmlm}, at a given self-similar abscissa $x/\lambda_2$,
shows that the MLM pressure 
tends to the linearised Bernoulli pressure
as ${\rm d} \lambda_2 / {\rm d} h \rightarrow +\infty $.

\vspace{-0.1cm}
\section{Vertical water entry of a foil}
\label{sec_naca}

\subsection{NACA foil: geometry, flow separation, and choice of continuation angles}
\label{subsec_naca_geom}

% $$$$$$$$ TABLE
 \begin{table}[t]
\begin{center}
\begin{tabular}{|c|c|c|c|c|c|c|}
\hline
& $w$ & $R_0 / c$ & $\delta \ [^{\circ}]$ & $\theta_s \ [^{\circ}]$ & $\theta_m \ [^{\circ}]$ & $R_m / c$ \\
\hline
NACA 0005 & $0.05$ & $2.76\cdot 10^{-3}$ & $3.35$ & $-0.71$ & $-2.59$ & $18.1$ \\
NACA 0010 & $0.1$ & $1.10\cdot 10^{-2}$ & $6.67$ & $-1.5$ & $-5.19$ & $9.14$ \\
NACA 0020 & $0.2$ & $4.41\cdot 10^{-2}$ & $13.2$ & $-3.2$ & $-10.4$ & $4.74$ \\
NACA 0028 & $0.28$ & $8.64\cdot 10^{-2}$ & $18.1$ & $-4.9$ & $-14.5$ & $3.55$ \\
NACA 0030 & $0.3$ & $9.92\cdot 10^{-2}$ & $19.3$ & $-5.5$ & $-15.5$ & $3.36$ \\
\hline
\end{tabular}
\end{center}
\caption{
Some properties of the NACA foils considered in the present study. 
$w$ is the maximum thickness, normalised by the chord length $c$.  
$R_0$ is the radius of curvature at the leading edge.
$\delta$ is the half-opening angle of the trailing edge.
$\theta_s$ is the inclination angle for which flow separation is simultaneous at the leading and 
trailing edges, within the FBC model.
$R_m$ is the maximum radius of curvature along the foil contour.
$\theta_m$ is the inclination angle for which the first fluid-body contact occurs at the point 
where the radius of curvature is maximum.
}
\label{tab_geometry}
\end{table}
% $$$$$$$$ TABLE

\begin{figure}[t]
\begin{center}
\begin{tabular}{>{\centering\arraybackslash}m{.45\textwidth} >{\centering\arraybackslash}m{.55\textwidth}}
        \includegraphics[width=0.45\textwidth]{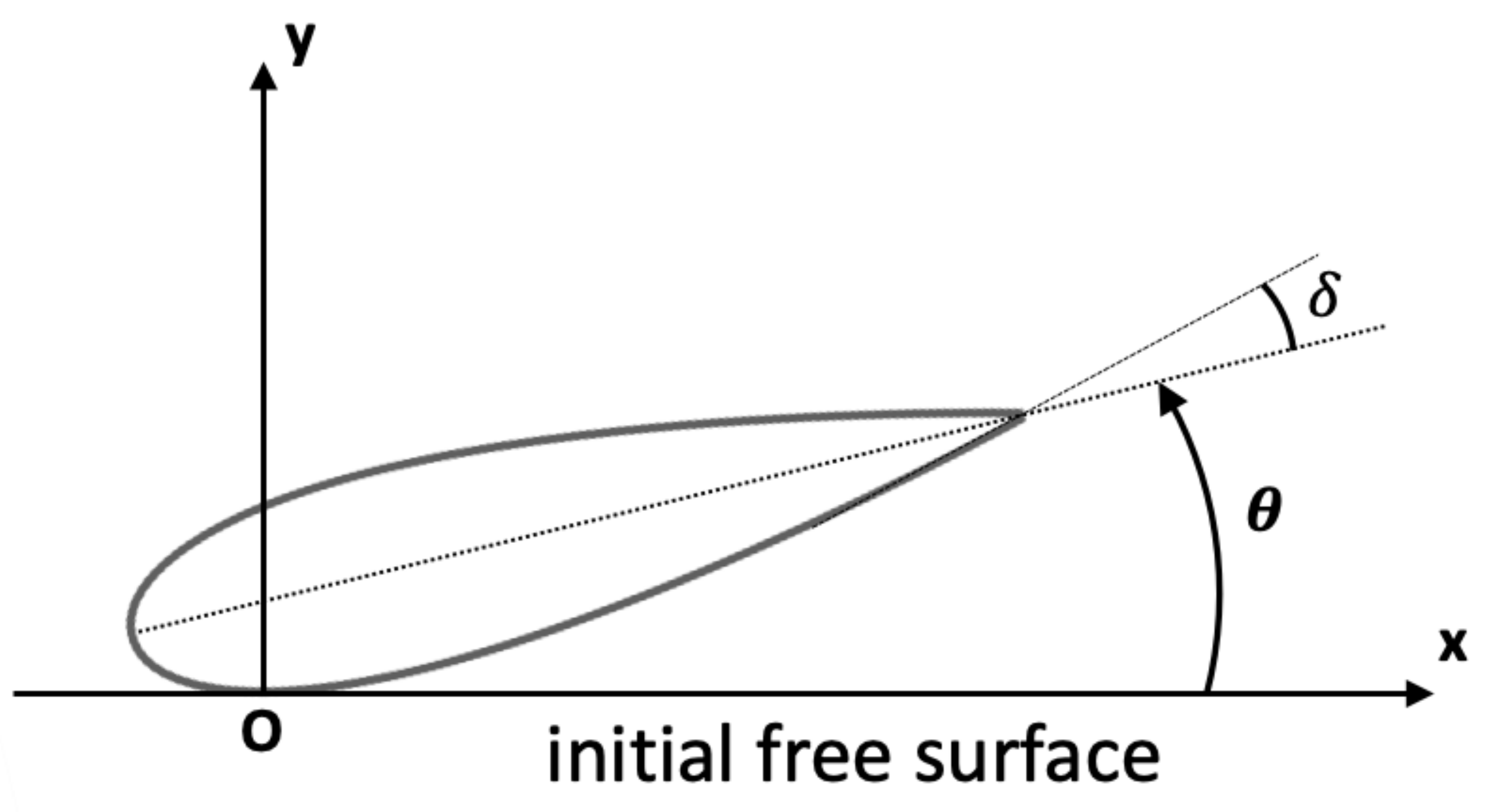}   &
        \includegraphics[width=0.55\textwidth]{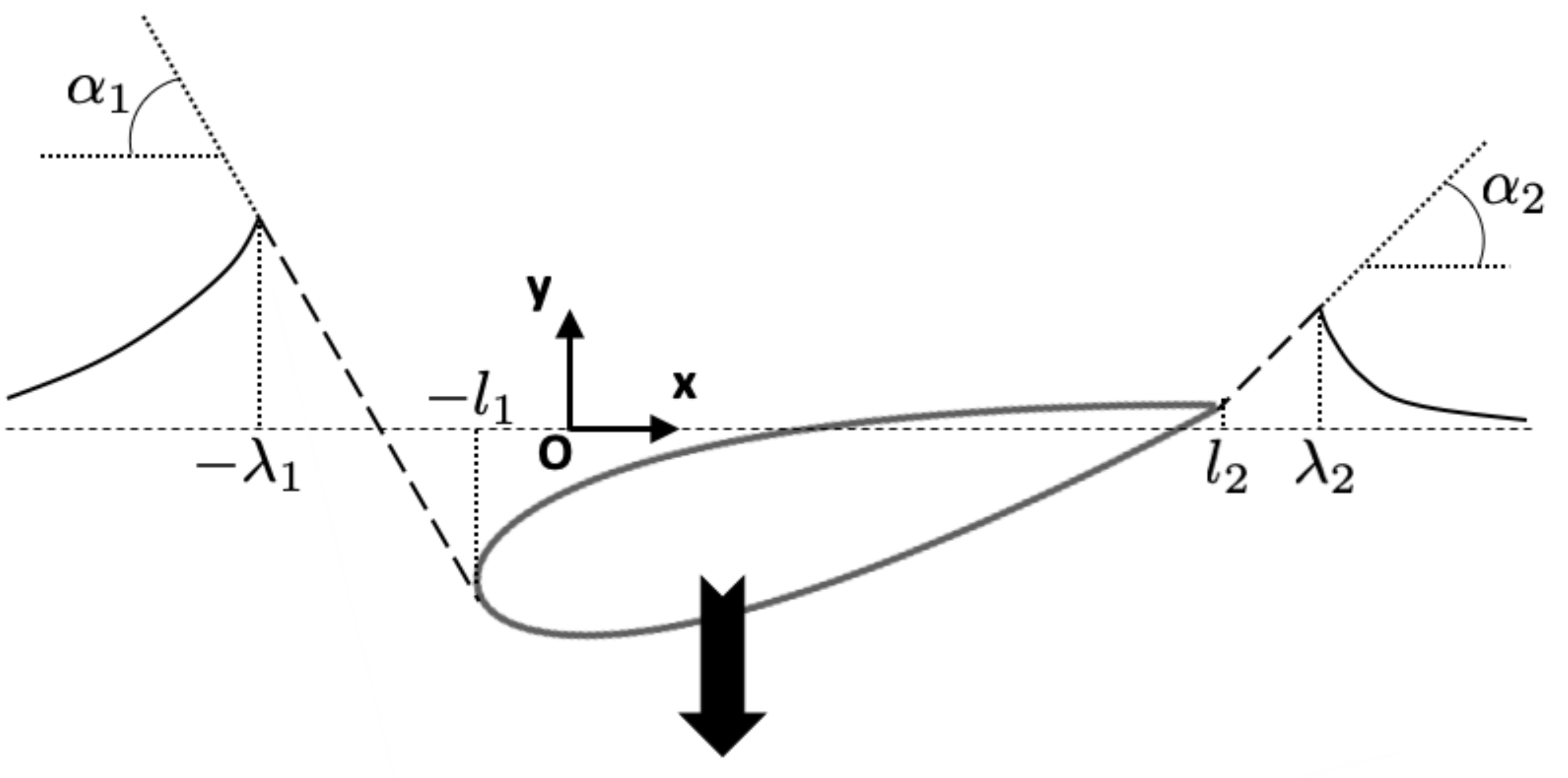} \\
        (a) & (b)
\end{tabular}
\end{center}
\caption{ 
(a): 
Initial conditions of impact when the foil first touches the water.
The initial free surface is assumed to be flat;
$\theta$ is the inclination angle of the foil with respect to the initial free surface; 
$\delta$ is the half opening angle of the trailing edge.
(b): Fictitious body continuation after flow separation from both sides of the foil.
The foil is continued by two flat plates (dashed lines) of inclinations, 
$\alpha_{1}$ on the left, and $\alpha_2$ on the right.
}
\label{fig_naca_sf}
\end{figure}

In this paragraph, we go one step further by considering the vertical water entry of a NACA\footnote{
The National Advisory Committee for Aeronautics (NACA) was a US federal agency,
replaced with the National Aeronautics and Space Administration (NASA) in 1958.} foil \cite{abbott_1945}.
The half-thickness of a symmetric NACA foil is given by 
\vspace{-0.1cm}
\begin{equation}
\label{eq_naca_profile}
\begin{split}
g(\xi) = c\cdot \frac{\thickfoil}{0.2} \left[ 
0.2969 \sqrt{\frac{\xi}{c}} 
- 0.1260 \left( \frac{\xi}{c} \right) 
- 0.3516 \left(\frac{\xi}{c}\right)^2  
+ 0.2843 \left(\frac{\xi}{c}\right)^3
- 0.1015 \left(\frac{\xi}{c}\right)^4 
\right] \, ,
\end{split}
\end{equation}
where $c$ is the chord length, 
$c\cdot \thickfoil$ is the maximum thickness and $\xi$ is the position along the chord 
($\xi=0$ at the leading edge, $\xi=c$ at the trailing edge).
The leading edge of the foil approximates a circular cylinder of radius
\begin{equation}
\label{eq_r0}
R_0 \simeq 1.102 \thickfoil^2 c \, ,
\end{equation}
and the half opening angle of the trailing edge is given by 
\begin{equation}
\label{eq_delta}
\delta = {\rm atan}(1.169 \ \thickfoil) \, .
\end{equation}
In \S\ref{subsect_cfd_compare}, we consider a NACA 0028 foil, whose contour follows \eq{eq_naca_profile} with $\thickfoil=0.28$
(see \fig{fig_naca_sf} for illustration). 
NACA foils with other thicknesses are considered in \S\ref{subsec_diff_thick}; 
some characteristics of the foils considered in the present paper are given in Tab. \ref{tab_geometry}.

% $$$$$$$$$$$$$$$$$$$$$$$$$$$$$$$$$$$$$$
% $$$$$$$$ FIGURES
% $$$$$$$$$$$$$$$$$$$$$$$$$$$$$$$$$$$$$$

% $$$$$$$$$$$$$$$$$$$$$$$$$$$$$$$$$$$$$$
% $$$$$$$$ FIGURES
% $$$$$$$$$$$$$$$$$$$$$$$$$$$$$$$$$$$$$$

During the water entry of a foil, 
flow separation can occur on both sides of the body,
requiring the choice of two continuation angles $\alpha_1$ and $\alpha_2$ within the FBC framework (see \fig{fig_naca_sf}).
At the foil leading edge,
a tangential connection between the real body and the fictitious 
sloped
plate is imposed;
thus, the choice of $\alpha_1$ also sets
the connection point between the fictitious 
plate and the foil leading edge.
Then, the flow separation takes place at the point of the body contour where the local deadrise angle is equal to $\alpha_1$.
This tangential connection is motivated by experiments and numerical results, 
which show that the flow separation from a smooth body leads to a smooth transition in terms of the hydrodynamic force.
In principle, the continuation angles $\alpha_1$ and $\alpha_2$ could be functions of $\theta$. 
In the present paper, we 
investigate to what extent the choice of  $\alpha_1$ and $\alpha_2$ can be generic,
and set the values,
only depending on the type of separation:
\begin{enumerate} 
\item At the leading edge, flow separation occurs on a smooth part of the contour.
In this case we set $\alpha_1 = 60^{\circ}$, which is the continuation angle obtained by Tassin et al. (2014) \cite{tassin_2014} for a circular cylinder.
\item Close to the trailing edge, the radius of curvature of the contour becomes large, and separation occurs at a chine.
Therefore we set $\alpha_2 = 47^{\circ}$, which is the continuation angle recommended by Tassin et al. (2014) \cite{tassin_2014} for a horizontal flat plate.
This choice is also consistent with the results obtained in Section \ref{sec_inclined_fp} for the water entry of an inclined flat plate.
\end{enumerate}

\subsection{CFD simulations}

In order to have a point of comparison and assess the applicability of the FBC concept to 
a foil shape,
the same water entry configurations were investigated by means of Computation Fluid Dynamics.
The simulations have been carried out with the finite-element software ABAQUS/Explicit (version 2017), 
using the coupled Eulerian-Lagrangian formulation. 
In this framework, 
the impacting body is modelled as a Lagrangian solid (rigid in the present case), 
while the fluid flow is described using the Eulerian approach. 
The position of the fluid surface is tracked using the Volume-Of-Fluid (VOF) method, 
coupled with an interface-reconstruction scheme. 
As viscosity does not play a significant role in water impact problems, it is not considered in the present simulations.
Accounting for viscosity would have required a very fine mesh, to properly resolve the boundary layer.
The purpose of these simulations is to validate the FBC concept, therefore gravity is not included.
The general contact algorithm of ABAQUS is used to describe the interaction between the fluid and the solid. 
In this method, only compressive stresses are transmitted across the fluid-solid interface, 
and flow separation occurs when the pressure on the body contour drops to zero.

As the ABAQUS/Explicit solver is not able to deal with incompressible flows, fluid compressibility was taken into account. 
For the impact of a \textit{blunt} body, compressibility matters only at the very beginning of water entry. 
At small penetration depths, a blunt body contour approximates a parabola and the expansion velocity of the Wagner wetted area follows
\begin{equation}
\dot{\lambda}_1 \simeq \dot{\lambda}_2 \simeq \sqrt{ \frac{R_c}{h}} \dot{h} \, , \ \ \, {\rm for} \ h\rightarrow 0 \, ,
\end{equation}
where $R_c$ is the radius of curvature of the body contour at the point of first contact with the fluid.
The typical penetration depth, over which fluid compressibility matters (see for example \cite{korobkin_1988}),
can be obtained by equating the expansion velocity of the Wagner wetted surface with the speed of sound in the liquid, $c_l$:
\begin{equation}
h_c = \left(\frac{ \dot{h}}{c_l}\right)^2 R_c \, .
\end{equation}
In the CFD simulations reported hereafter, 
the impacting solid is a NACA 0028 foil with a chord length $c=1$ m 
and a velocity at first contact with water, $\dot{h} = 1 \ {\rm m/s}$ (except for one simulation reported in \S\ref{par_abaqus_acc},
where the solid body starts at rest).
The speed of sound is set to $c_l=500 \ {\rm m/s}$
and the maximum radius of curvature along the foil contour is $3.5$ m, leading to $h_c \simeq 1.4 \cdot 10^{-5}$ m. 
This value is much smaller than the penetration depths relevant to slamming loads (see Figs. \ref{fig_naca_vert_1}-\ref{fig_naca_vert_2}).

\begin{figure}[h]
\begin{center}
\begin{tabular}{c}
        \includegraphics[width=0.5\textwidth]{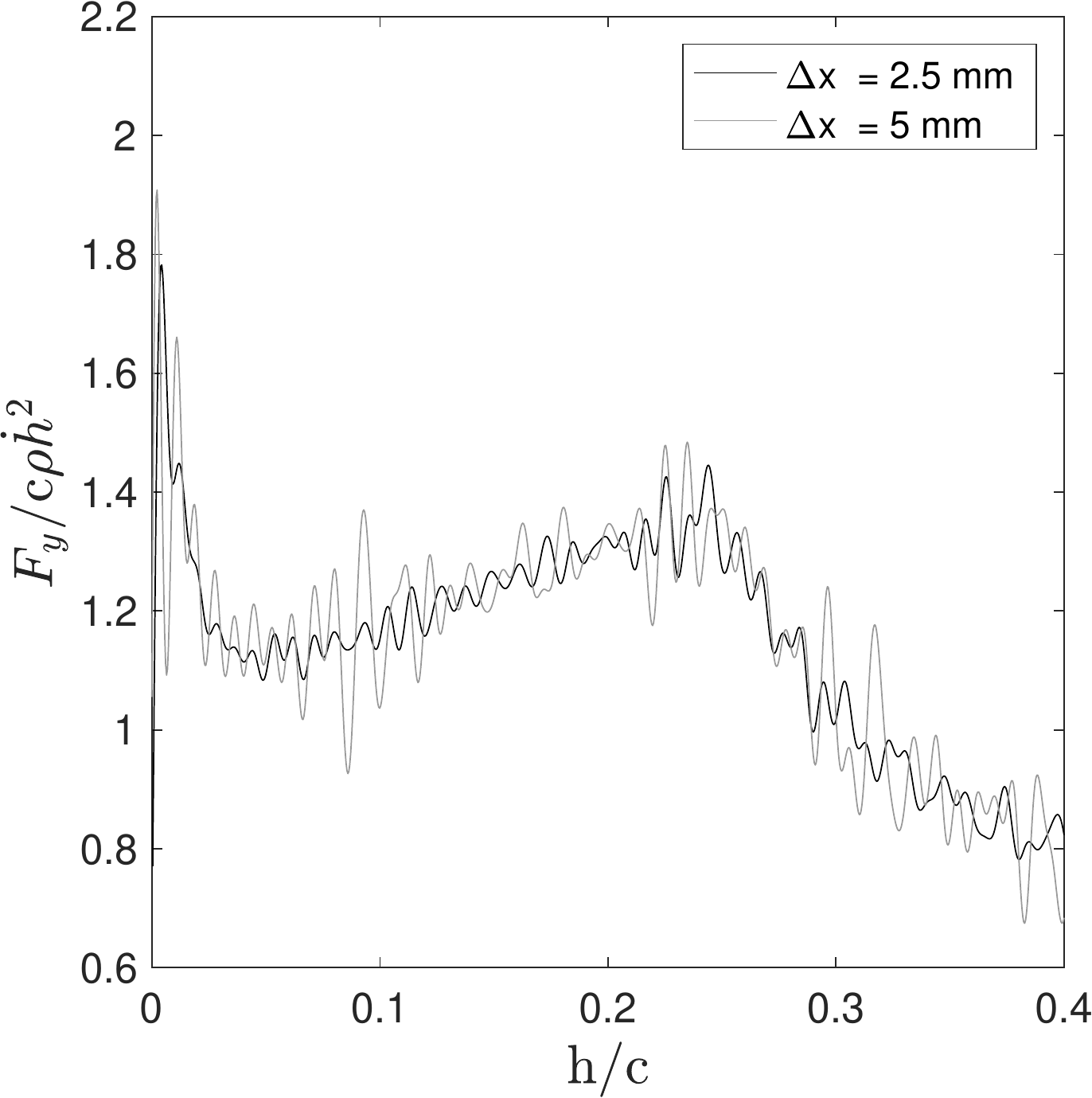}
\end{tabular}
\end{center}
\caption{
The effect of mesh resolution on ABAQUS simulations.
The quantity shown is the nondimensional vertical force component as a function of the nondimensional penetration depth, 
for the water entry of a NACA 0028 foil at constant velocity $\dot{h} = 1$ m/s, with an inclination angle $\theta = 20^{\circ}$.
CFD results are shown for two different grid spacings: $\Delta x = 5$ mm (grey curve) and $\Delta x = 2.5$ mm (black curve).
}
\label{fig_abaqus_mesh}
\end{figure}

The Eulerian fluid domain is $20$ m in width and $10$ m in height. 
Non-reflecting boundary conditions are 
enforced
at the remote 
boundaries 
of the fluid domain
to minimise the reflection of acoustic waves.
Two different meshes have been considered. 
The first mesh features 347733 elements with a mesh spacing of $\Delta x = 5$ mm in the region where the fluid-solid interaction takes place. 
The second mesh features 703800 elements with a mesh spacing in the impact area of $\Delta x = 2.5$ mm. 
\fig{fig_abaqus_mesh} shows the evolution of the vertical component of the hydrodynamic force obtained with the two meshes, 
for the vertical water entry of a NACA 0028 foil at constant velocity.
Both curves show some high-frequency oscillations. 
These oscillations are 
probably due to the Euler-Lagrange contact algorithm, which is based on a penalty method \cite{aquelet_2006}. 
In a previous study \cite{tassin_2012}, the present CFD solver has been shown to provide accurate estimates of slamming loads
for various body shapes.
When the mesh is refined, the magnitude of the oscillations is reduced. However, the overall force evolution remains unchanged. 
All results presented in the rest of the present paper have been obtained with the finest mesh ($\Delta x = 2.5$ mm).

\subsection{Comparison of the analytical model with CFD results}
\label{subsect_cfd_compare}

Figs. \ref{fig_naca_vert_1}-\ref{fig_naca_vert_2} show the evolution of 
the two hydrodynamic force components $F_x$, $F_y$, 
and of the 
moment $M_z$ 
(expressed at the foil leading edge) 
acting on a NACA 0028 foil, during vertical water entry at constant velocity.
The FBC predictions are compared with the CFD results for 5 different inclination angles: 
$\theta  = -28.1^{\circ}$; $-18.1^{\circ}$; $-14.5^{\circ}$; $0^{\circ}$; $20^{\circ}$.
\fig{fig_naca_acc} shows two additional comparisons for water entries at constant acceleration.

The results reported in this section are nondimensionalized 
and given in terms of hydrodynamic coefficients. 
Semi-analytical results were directly computed in a dimensionless form.
Regarding CFD simulations, for the water entries at constant velocity, 
a specific value of the velocity, $\dot{h} = 1$ m/s, had to be specified. 
As long as the fluid viscosity, surface tension, flow compressibility, and gravity effect can be neglected, 
the chosen value of the velocity is expected to have no effect on the 
slamming 
coefficients reported in Figs. \ref{fig_naca_vert_1}-\ref{fig_naca_vert_2}.
For the water entries at constant acceleration, 
an acceleration, $\ddot{h}$, and an initial velocity, $\dot{h}(0)$, had to be specified.
In these cases, an added-mass load adds to the slamming load (see Eq. \ref{eq_pmlm}).
A rescaling of the chord length 
and body kinematics keeping the ratio $\dot{h}(0)^2 / c \, \ddot{h}$ unchanged 
would have no effect on the hydrodynamic coefficients reported in \fig{fig_naca_acc},
as long as the same above-mentioned assumptions are valid.
In the following paragraphs, we discuss several aspects of our results.

% $$$$$$$$$$$$$$$$$$$$$$$$$$$$$$$$$$$$$$
% $$$$$$$$ FIGURES
% $$$$$$$$$$$$$$$$$$$$$$$$$$$$$$$$$$$$$$

\newcommand\figHeight{0.5}
\newcommand\figCropLeft{0.07}
\newcommand\figCropRight{0}

\begin{figure}[t]
\begin{center}
\begin{tabular}{ccc}
        \includegraphics[height=\figHeight\textheight]{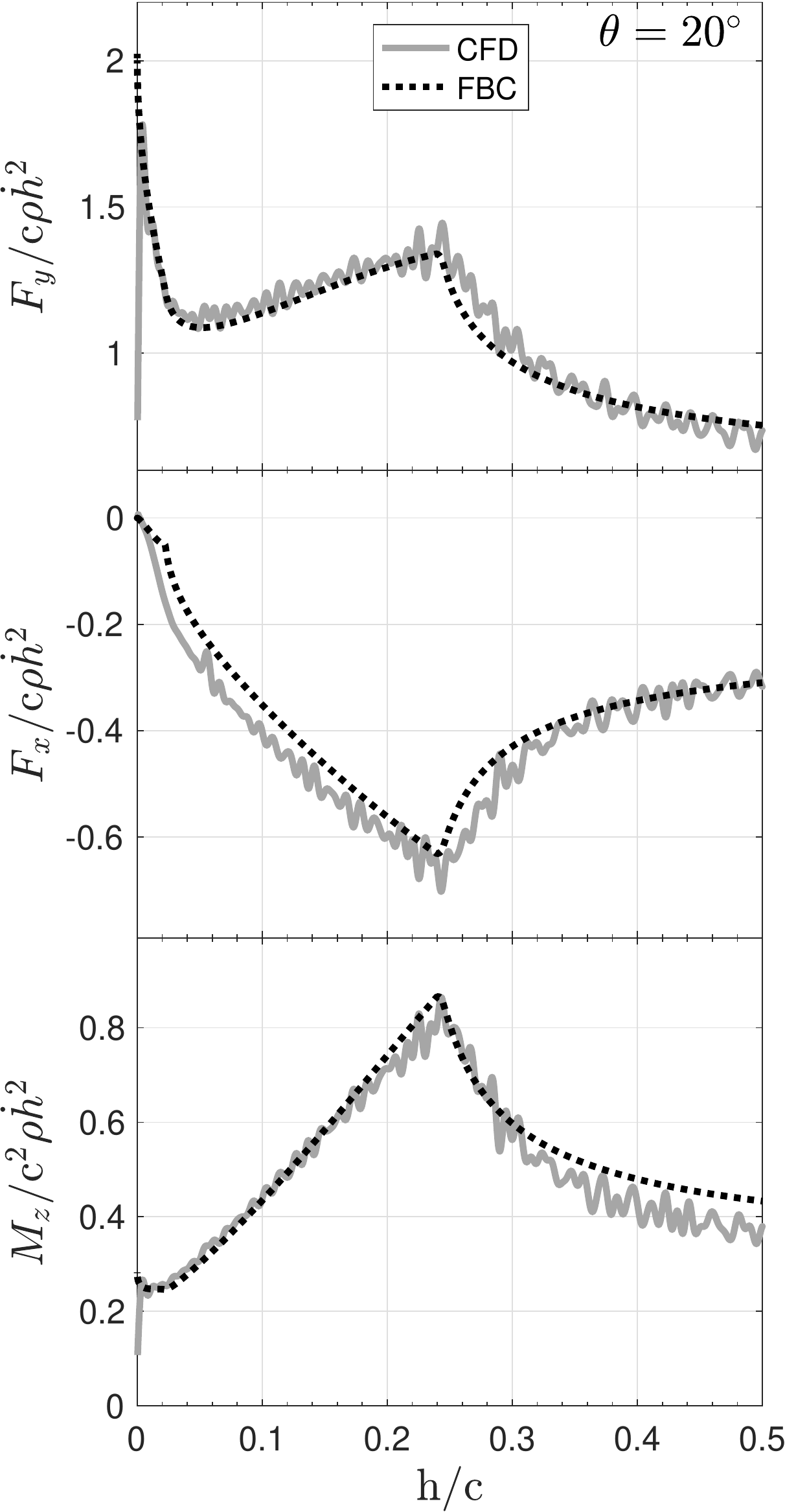}  & 
        \adjincludegraphics[height=\figHeight\textheight, trim={{\figCropLeft\width} 0 {\figCropRight\width} 0},clip]{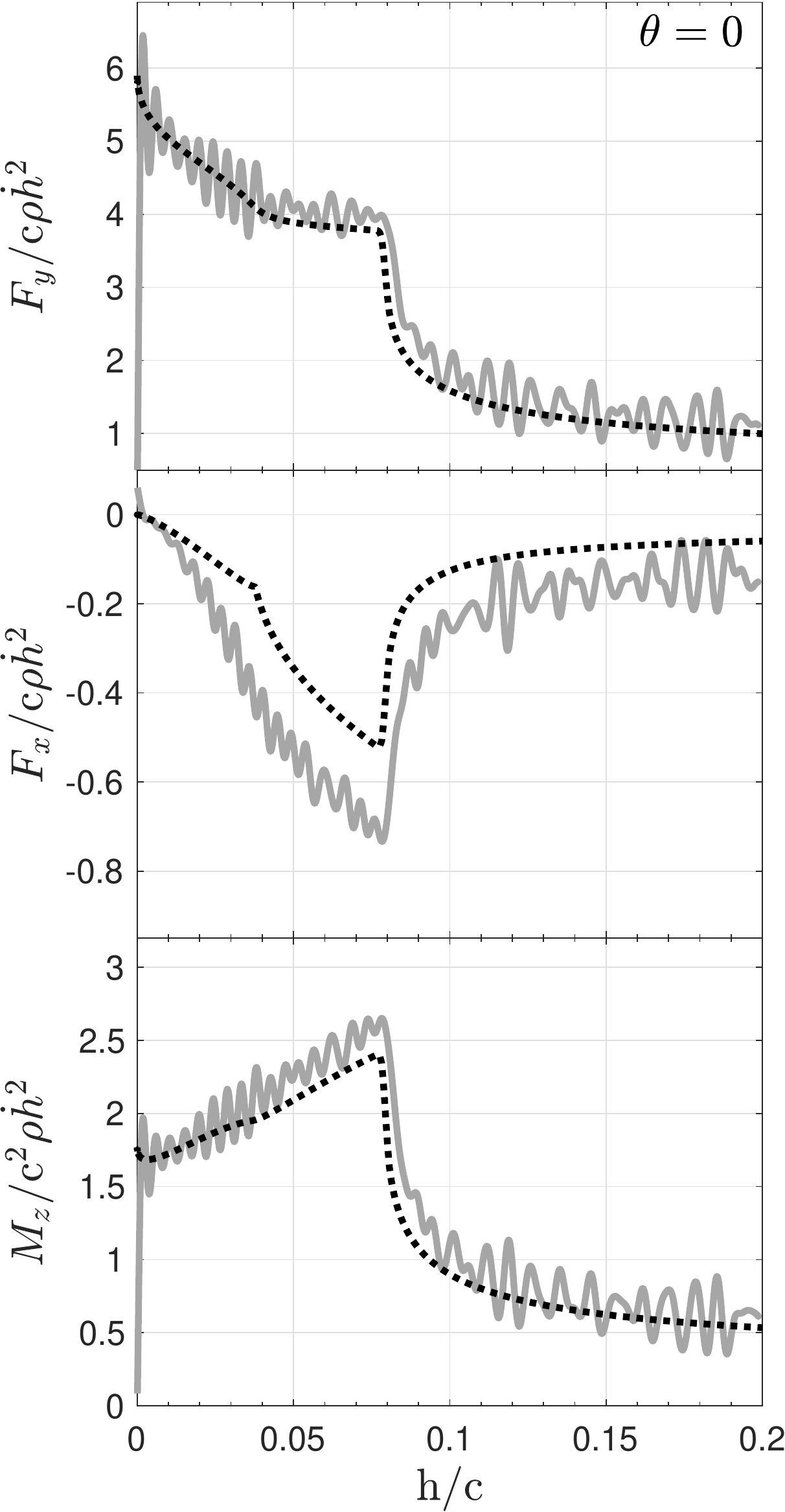} &  
        \adjincludegraphics[height=\figHeight\textheight, trim={{\figCropLeft\width} 0 {\figCropRight\width} 0},clip]{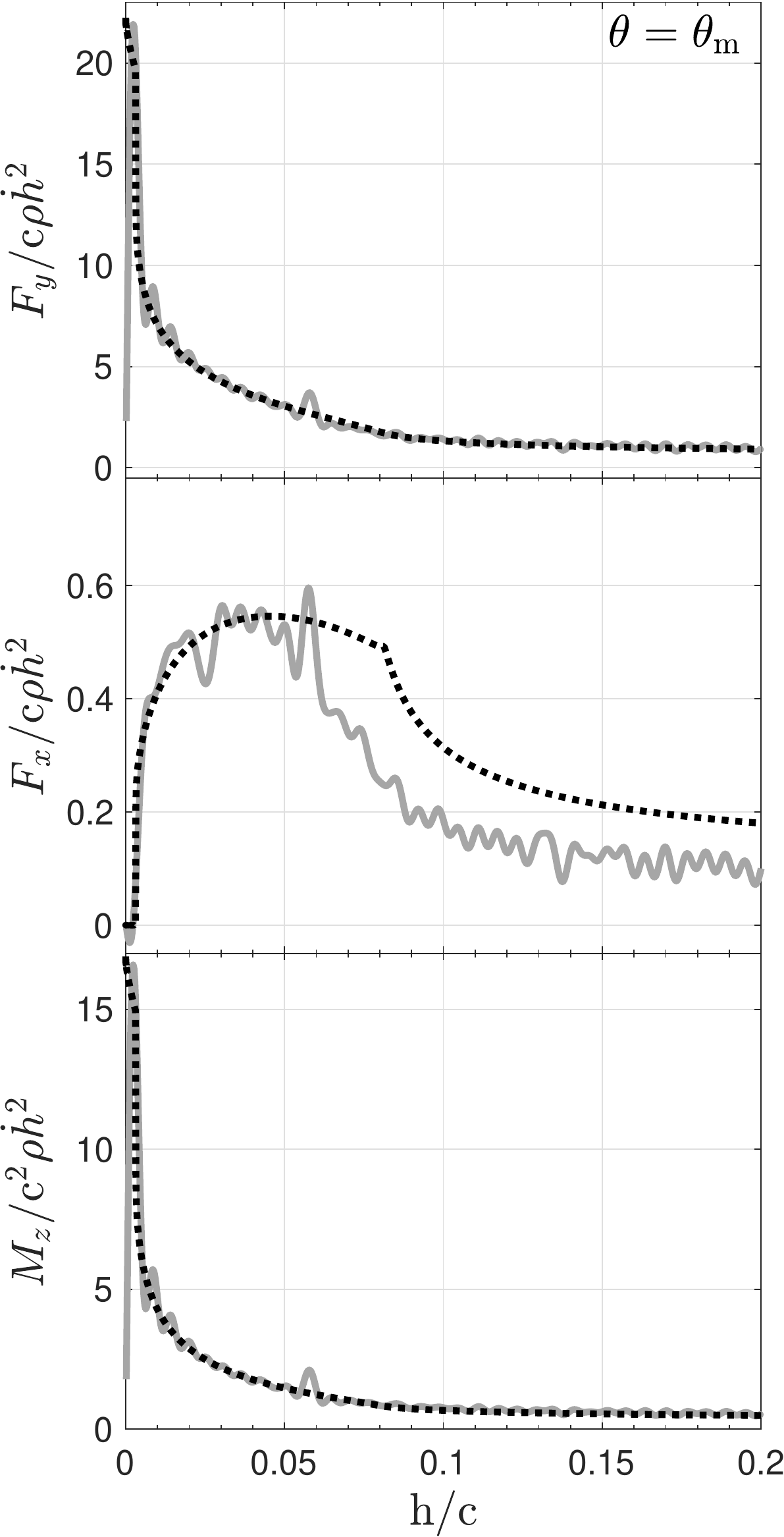}    
\end{tabular}
\end{center}
\caption{
Vertical water entry of a NACA 0028 foil at constant velocity.
CFD and FBC results are shown respectively as grey solid lines and black dotted lines.
From top to bottom: normalised force components and moment, $F_y$, $F_x$, $M_z$.
From left to right: calculations are shown for different inclination angles, $\theta = 20^{\circ}$ (left), $\theta = 0$ (middle), $\theta = \theta_m$ (right).
$\theta_{\rm m} = -14.5^{\circ}$ is the inclination angle at which the FBC model predicts maximum instant value for $F_y / c \rho \dot{h}^2$
(at the beginning of the impact).
}
\label{fig_naca_vert_1}
\end{figure}

\begin{figure}[t]
\begin{center}
\begin{tabular}{cc}
        \includegraphics[height=\figHeight\textheight]{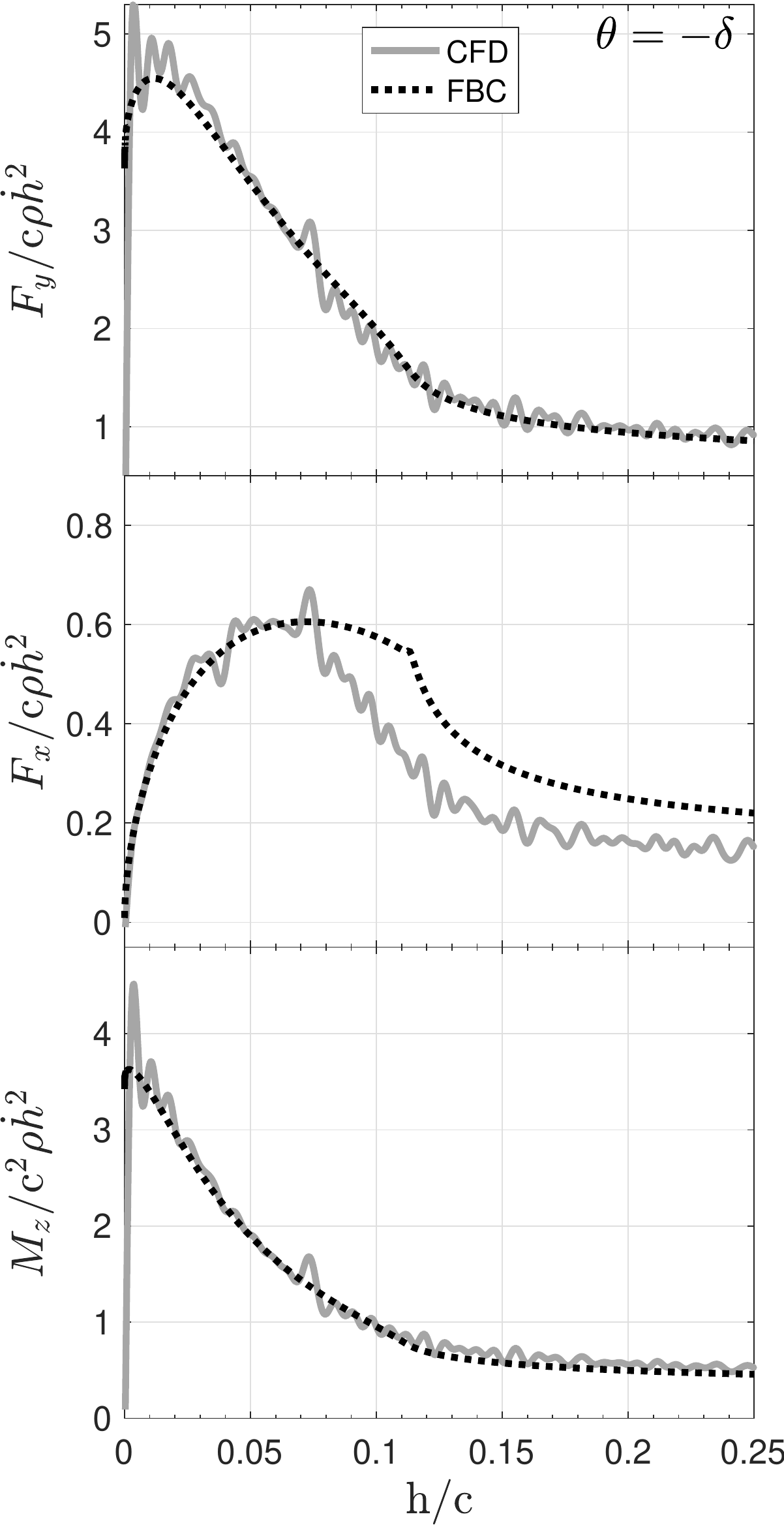}  & 
        \adjincludegraphics[height=\figHeight\textheight, trim={{\figCropLeft\width} 0 {\figCropRight\width} 0},clip]{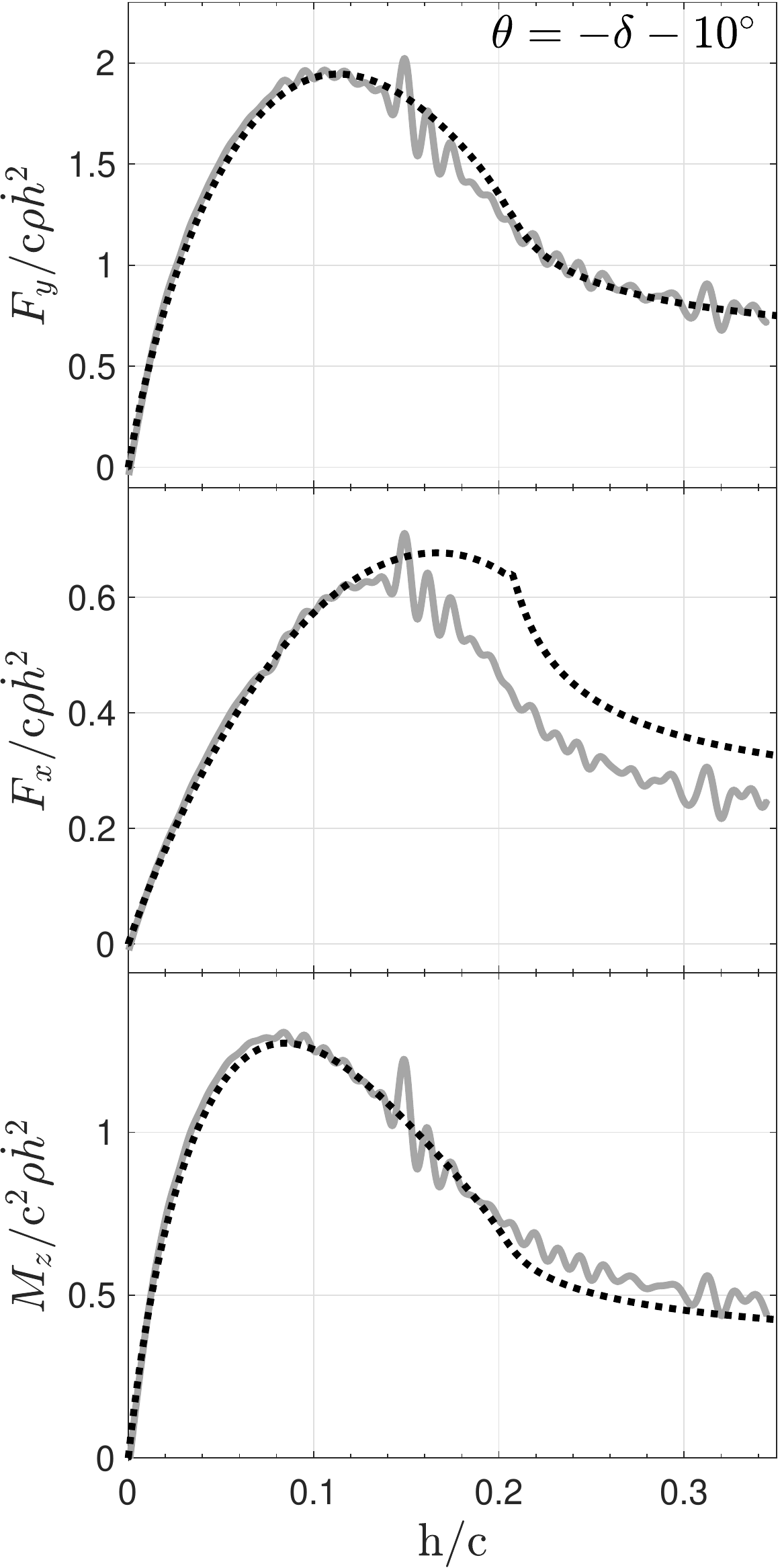}     
\end{tabular}
\end{center}
\caption{
Same as \fig{fig_naca_vert_1} for two other inclination angles: $\theta = -\delta $ (left) and $\theta = -\delta-10^{\circ}$ (right).
$\delta \simeq 18.1^{\circ}$ is the half opening angle of the foil trailing edge (see Eq. \ref{eq_delta}).
For $\theta = -\delta $ the trailing edge contour is tangent to the initial free surface.
}
\label{fig_naca_vert_2}
\end{figure}

% $$$$$$$$$$$$$$$$$$$$$$$$$$$$$$$$$$$$$$
% $$$$$$$$ FIGURES
% $$$$$$$$$$$$$$$$$$$$$$$$$$$$$$$$$$$$$$

\subsubsection{Times of flow separation} 
For $\theta  = 0^{\circ}; 20^{\circ}$, 
the flow separation first occurs at the leading edge and then at the trailing edge.
For $\theta = -28.1^{\circ}; -18.1^{\circ}; -14.5^{\circ}$, the flow separation sequence is reversed. 
For $\theta = -28.1^{\circ}; -18.1^{\circ}$, the flow separates from the trailing edge right at the first fluid-body contact.
Flow separation events can be easily identified for the FBC model as they 
induce discontinuities in the slope of the force-displacement curves.
Flow separation transitions can also be identified in CFD results at the same penetration depths,
although they are somewhat smoothened in some of the $F_x$ curves.
For all inclination angles, both models show a good agreement on separation times.

\subsubsection{Force components $F_x$, $F_y$ and moment $M_z$}
The FBC and CFD models agree 
very well in terms of 
vertical force $F_y$ and moment $M_z$,
for all considered inclination angles. 
The agreement is less satisfactory regarding the horizontal force component $F_x$ (except for $\theta=20^{\circ}$).
We note, however, that the magnitude of $F_x$ is significantly smaller than the magnitude of $F_y$ for the considered range of inclination angles.
This larger disagreement is not surprising since $F_x$, given by (see \S\ref{subsec_fbc_cont} for an explanation about the limits of integration)
\begin{equation}
\label{eq_fx}
F_x(t) = - \int_{-{\rm min}(\lambda_1,l_1)}^{{\rm min}(\lambda_2,l_2)} P(x,t) f_{,x}(x) {\rm d} x \, ,
\end{equation}
is a second-order quantity
whose main contributions are concentrated close to the contact points.
Indeed, the splash root is the region where the pressure \textit{and} the deadrise angles (i.e. $f_{,x}$) are the largest for a curved contour.
Thus, the integral in \eq{eq_fx}
is largely dominated by regions where the flow nonlinearities, ignored in Wagner's approach, are the strongest.
This explains why the MLM estimates for $F_x$ are not as reliable as for $F_y$ and $M_z$.
For practical use, it is not an issue as long as $F_x$ remains moderately smaller that $F_y$.
This last condition is guaranteed for body contours with moderate deadrise angles ($\la 30^{\circ}$), or body contours which are not strongly asymmetric.

\subsubsection{Water entry with acceleration}
\label{par_abaqus_acc}

% $$$$$$$$$$$$$$$$$$$$$$$$$$$$$$$$$$$$$$
% $$$$$$$$ FIGURE
% $$$$$$$$$$$$$$$$$$$$$$$$$$$$$$$$$$$$$$

\begin{figure}[t!]
\begin{center}
\begin{tabular}{c:c}
        $\dot{h}(0) = 1 \ {\rm  m/s}$, $\ddot{h}(t) = 5 \ {\rm  m/s^2}$ &
        $\dot{h}(0) = 0$, $\ddot{h}(t) = 5 \ {\rm  m/s^2}$ \\
        \includegraphics[height=\figHeight\textheight]{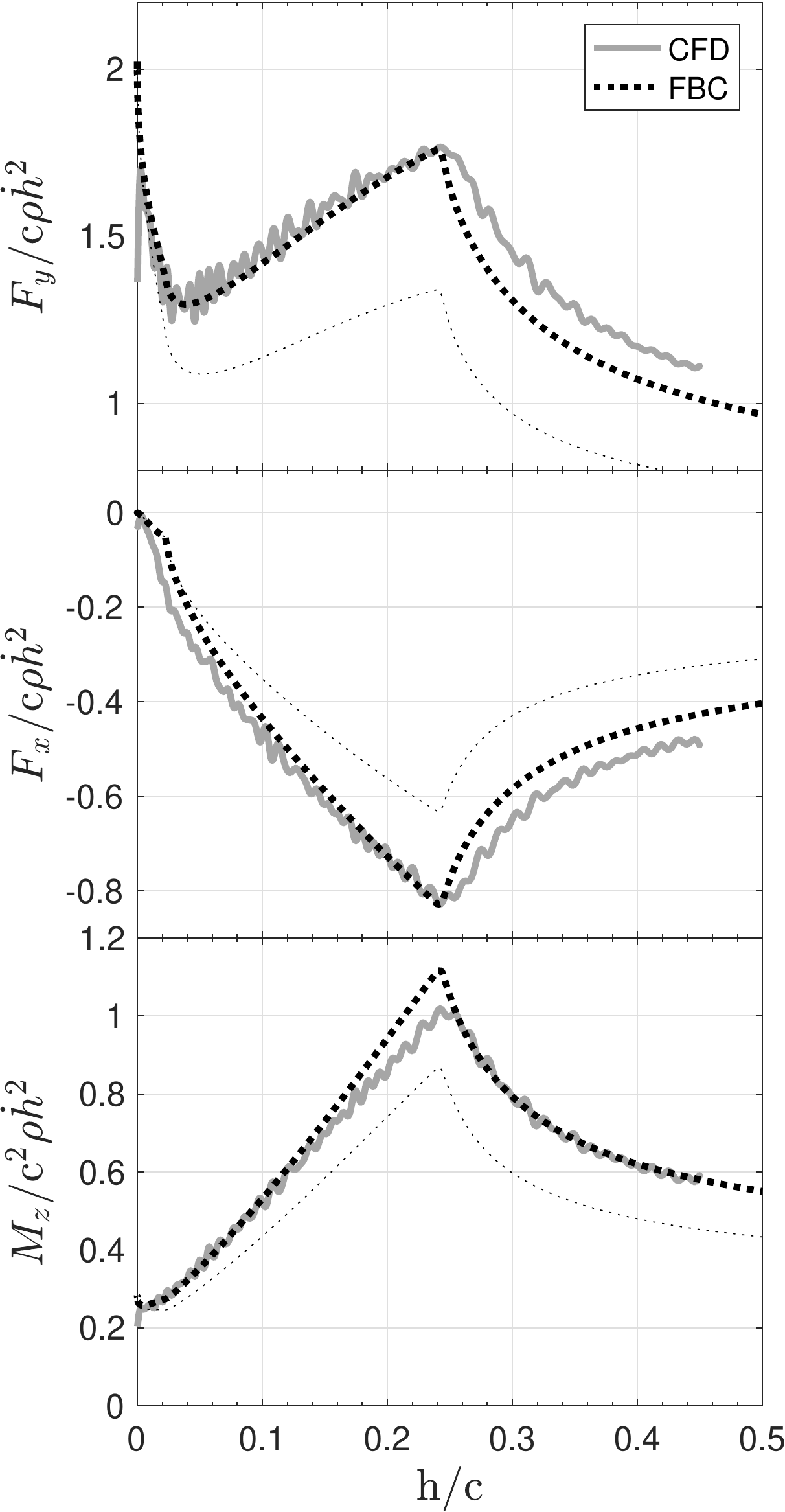}  & 
        \adjincludegraphics[height=\figHeight\textheight, trim={{\figCropLeft\width} 0 {\figCropRight\width} 0},clip]{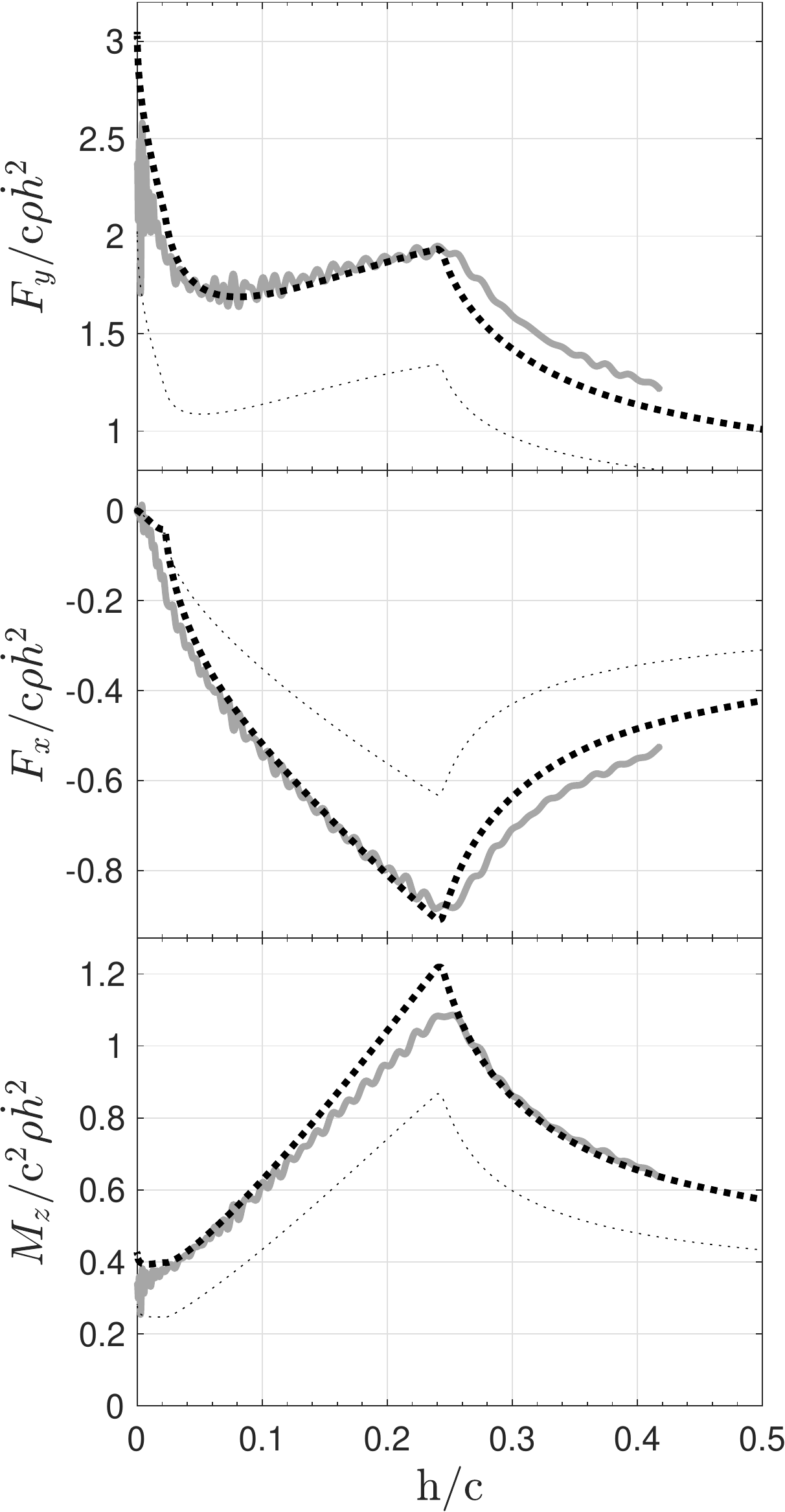}   
\end{tabular}
\end{center}
\caption{
Vertical water entry of a NACA 0028 foil (chord length, $c=1$ m) at constant acceleration, $\ddot{h}(t) = 5 \ {\rm m/s^2}$. 
CFD and FBC results are shown respectively as grey lines and black dotted lines.
From top to bottom: normalised force components and moment, $F_y$, $F_x$, $M_z$.
Calculations are shown for 2 different initial velocities (at first contact with the fluid): 
$\dot{h}(0) = 1 \ {\rm m/s}$ (left) and $\dot{h}(0) =0$ (right).
For comparison purpose, FBC results with no acceleration (same curves as in \fig{fig_naca_vert_1}, left panel) are shown as thin dotted lines.
}
\label{fig_naca_acc}
\end{figure}

% $$$$$$$$$$$$$$$$$$$$$$$$$$$$$$$$$$$$$$
% $$$$$$$$ FIGURE
% $$$$$$$$$$$$$$$$$$$$$$$$$$$$$$$$$$$$$$

\fig{fig_naca_acc} shows two examples of the simulated slamming load evolution when the foil impacts water at constant acceleration.
The foil is inclined at an angle $\theta = 20^{\circ}$ and is accelerated into the fluid at $\ddot{h}(t) = 5 \ {\rm m/s^2}$.
At first contact with the fluid two different initial velocities are considered: $\dot{h}(0) = 0$ and $\dot{h}(0) = 1 \ {\rm m/s}$.
The results obtained with the FBC approach for constant velocity are also depicted in \fig{fig_naca_acc} for the sake of comparison.
As the velocity used for the normalisation 
of $F_y$ is the \textit{instantaneous} velocity, 
the comparison with the curves obtained for constant velocity 
directly reflects the contribution of the added-mass force.
The FBC model predicts a slightly faster decay of the force after flow separation from the trailing edge
and a larger peak value of $M_z$ (by $\simeq10 \%$), compared to the CFD results.
Still, the FBC model gives a good estimate of the
added mass loads during the early stage of flow separation.

% $$$$$$$$$$$$$$$$$$$$$$$$$$$$$$$$$$$$$$
% $$$$$$$$ FIGURE
% $$$$$$$$$$$$$$$$$$$$$$$$$$$$$$$$$$$$$$

\begin{figure}[t!]
\begin{center}
\begin{tabular}{c}
        \adjincludegraphics[width=\textwidth]{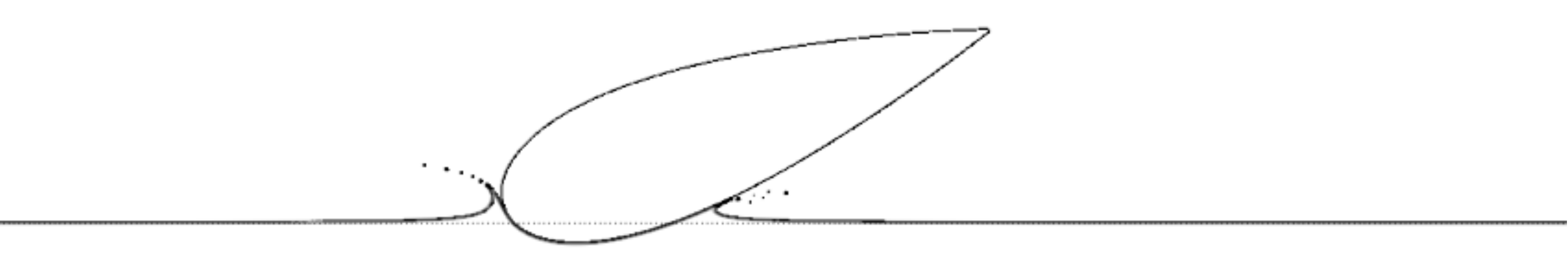}  \\
        	(a) before flow separation \\
        \adjincludegraphics[width=\textwidth]{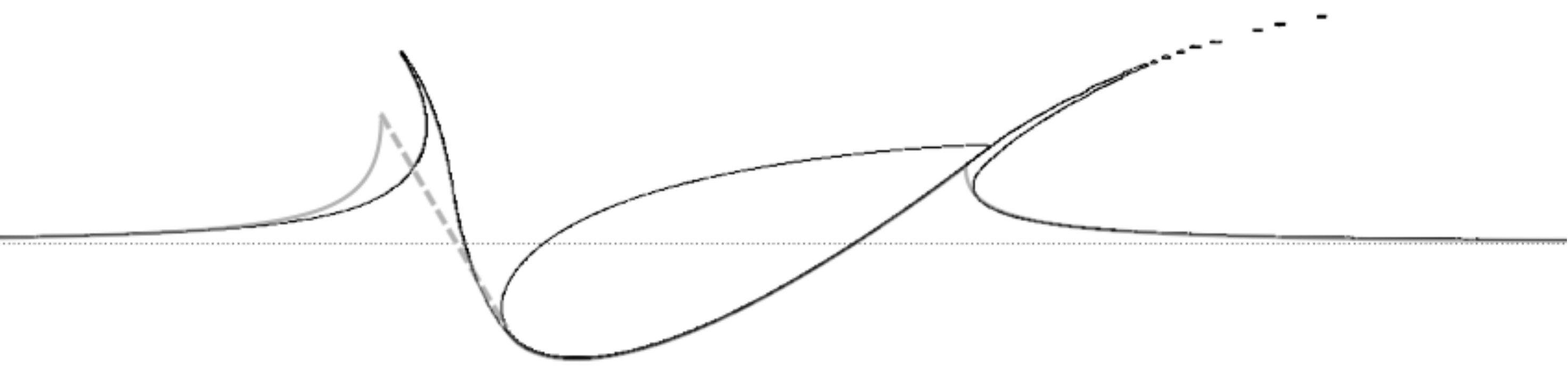}  \\
        (b) after flow separation from the leading edge \\
        \\
        \adjincludegraphics[width=\textwidth]{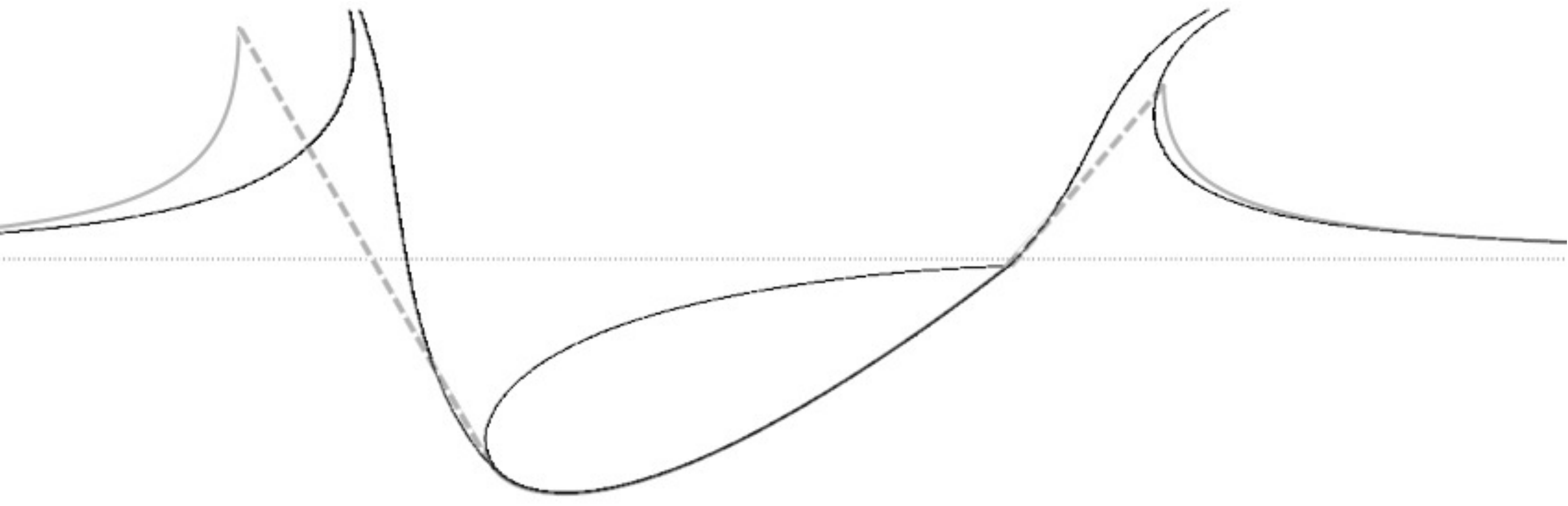}  \\
        (c) after flow separation from the trailing edge       
\end{tabular}
\end{center}
\caption{
Free surface of the flow induced by the vertical water entry of a NACA 0028 foil.
The inclination angle is $\theta=20^{\circ}$ and the entry velocity is constant, $\dot{h} = 1 \ {\rm m/s}$.
The free surface extracted from CFD simulations (black line) 
is compared with the fictitious flat plates (dashed grey lines) and the free surface (solid grey lines) obtained from the FBC model.
The initial free surface is shown as a thin dotted line.
Snapshots are given for 3 different penetration depths: (a) $h/c = 0.041$, (b) $h/c=0.23$, (c) $h/c=0.43$.
}
\label{fig_free_surface}
\end{figure}

% $$$$$$$$$$$$$$$$$$$$$$$$$$$$$$$$$$$$$$
% $$$$$$$$ FIGURE
% $$$$$$$$$$$$$$$$$$$$$$$$$$$$$$$$$$$$$$

\subsubsection{Free surface}
\label{subsubsec_abaqus_fs}

\fig{fig_free_surface} shows three snapshots of the free surface, 
taken during the water entry of the foil inclined at an angle $\theta=20^{\circ}$, with constant velocity.
The free surface profile extracted from the CFD simulations (black line) is compared 
with the fictitious flat plates and free surface elevation obtained from the FBC model (grey line).
Before flow separation (Fig.\ref{fig_free_surface}-a), the agreement between the CFD and Wagner models is excellent.

After flow separation from the leading edge, the FBC free surface rapidly 
deviates from the CFD free surface (Fig.\ref{fig_free_surface}-b).
However, on the other side of the body (trailing edge), the flow has not separated yet.
One can note the prominent water jet that has formed in the CFD simulation, and which is disregarded in Wagner's model.
As mentioned in the introduction, the splash jets do not contribute significantly to the hydrodynamic loads.
Except for the jet, both models show a good agreement on the expansion rate of the wetted surface towards the trailing edge.
This explains why the FBC model still provides satisfactory load estimates after flow separation from the leading edge.

After flow separation from both sides of the foil (Fig.\ref{fig_free_surface}-c), 
the FBC free surface elevation on the trailing edge side
also starts deviating from the CFD shape. 
Then it becomes more difficult to qualitatively explain 
why the agreement on predicted hydrodynamic loads remains good (see \fig{fig_naca_vert_1}).

% $$$$$$$$$$$$$$$$$$$$$$$$$$$$$$$$$$$$$$
% $$$$$$$$ FIGURES
% $$$$$$$$$$$$$$$$$$$$$$$$$$$$$$$$$$$$$$

\begin{figure}[t!]
\begin{center}
\begin{tabular}{cc}
        \includegraphics[height=\figHeight\textheight]{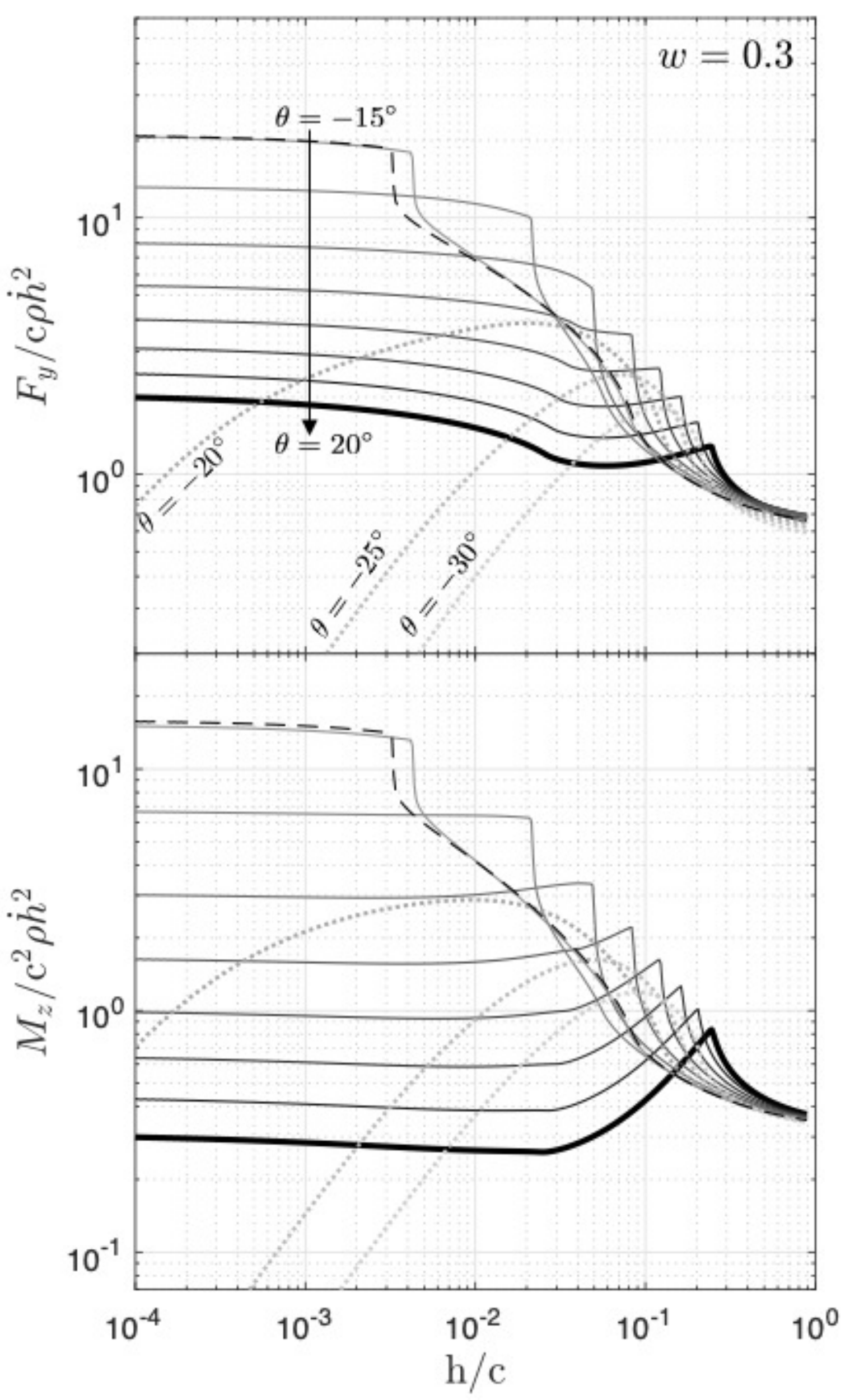}  & 
        \adjincludegraphics[height=\figHeight\textheight, trim={{\figCropLeft\width} 0 {\figCropRight\width} 0},clip]{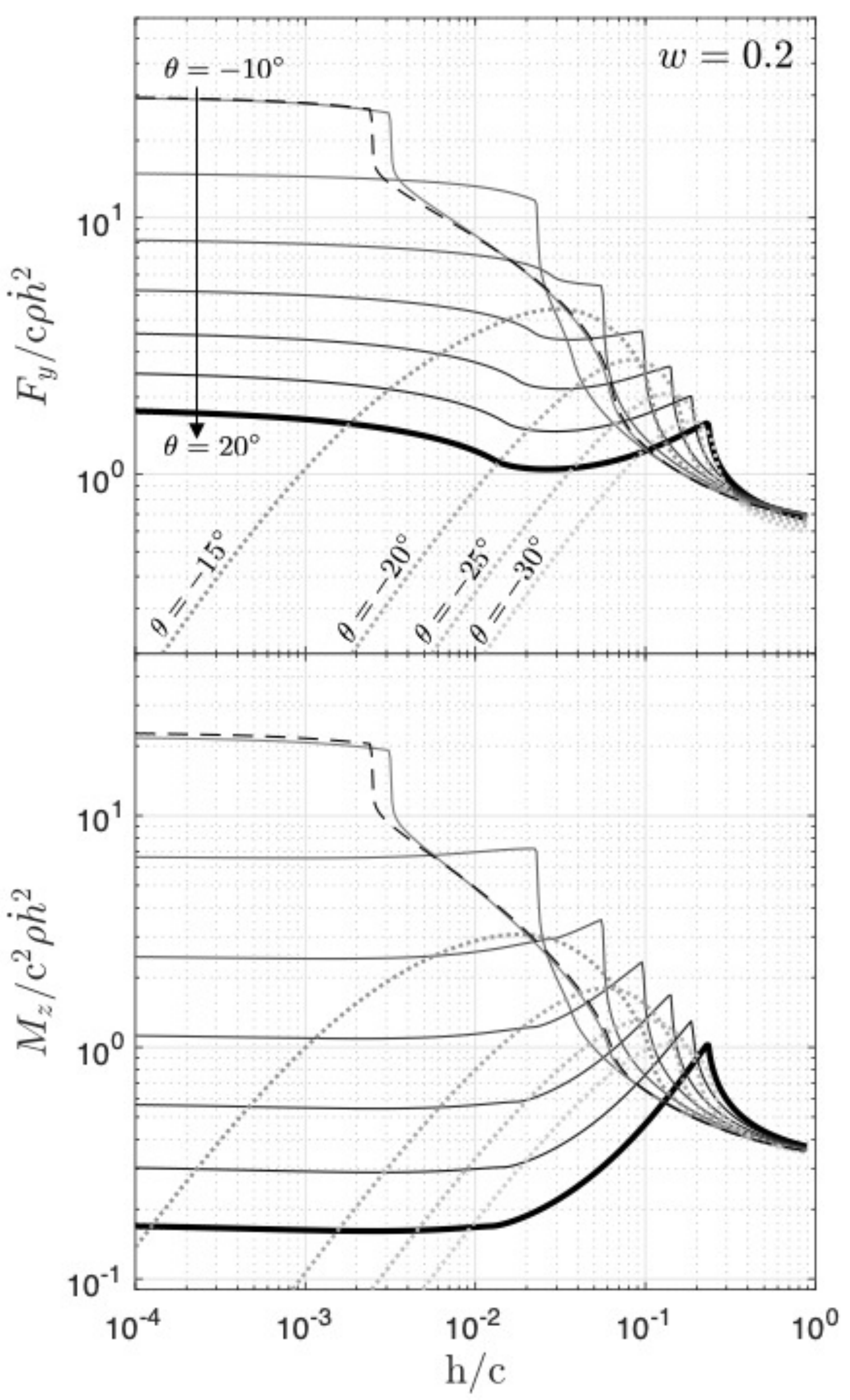}     
\end{tabular}
\end{center}
\caption{
Slamming loads on foils during a vertical water entry at constant velocity.
The evolution of the nondimensional vertical force component $F_y/c\rho \dot{h}^2$
and the nondimensional moment $M_z/c^2\rho \dot{h}^2$ (computed at the foil leading edge)
are given as a function of the nondimensional penetration depth $h/c$.
Results are shown for two relative foil thicknesses (see Eq. \ref{eq_naca_profile}): $w=0.3$ (left) and $w=0.2$ (right).
The slamming loads are plotted for different inclination angles using a gradation of grey, 
from $\theta = -30^{\circ}$ (light grey) to $20^{\circ}$ (black) with a $5^{\circ}$-step between two successive curves.
To ease the reading, 
dotted lines are used when the first contact with the fluid occurs at the trailing edge ($\theta<-\delta$, see Tab. \ref{fig_naca_sf}),
and the curve for $\theta = 20^{\circ}$ is thickened.
The dashed black lines show the slamming loads obtained for the inclination angle $\theta_m$ (see Tab. \ref{fig_naca_sf}), 
which gives the maximum instant loads (at first contact with the fluid).
}
\label{fig_naca_series_1}
\end{figure}

% $$$$$$$$$$$$$$$$$$$$$$$$$$$$$$$$$$$$$$
% $$$$$$$$ FIGURES
% $$$$$$$$$$$$$$$$$$$$$$$$$$$$$$$$$$$$$$

\begin{figure}[t!]
\begin{center}
\begin{tabular}{cc}
        \includegraphics[height=\figHeight\textheight]{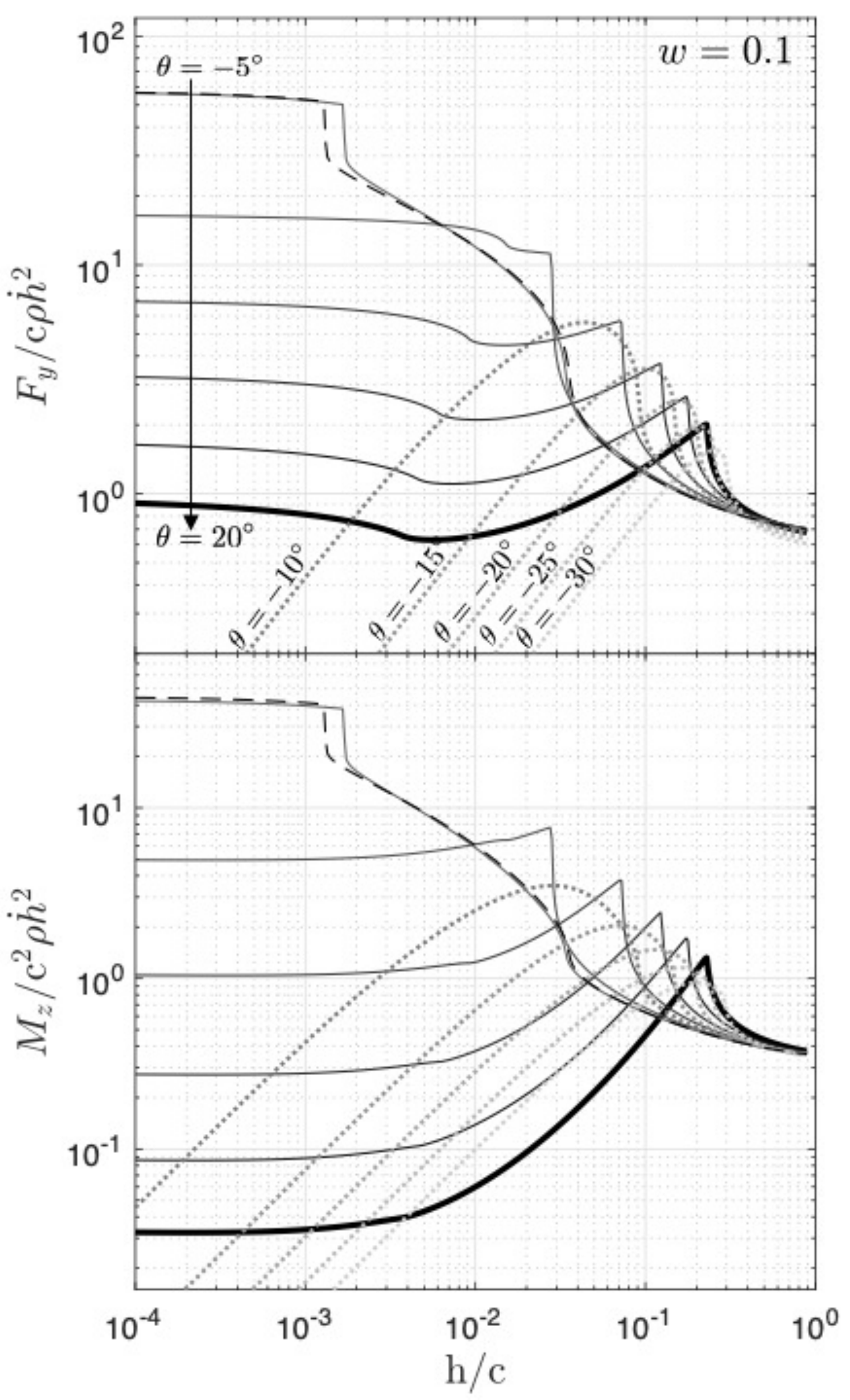}  & 
        \adjincludegraphics[height=\figHeight\textheight, trim={{\figCropLeft\width} 0 {\figCropRight\width} 0},clip]{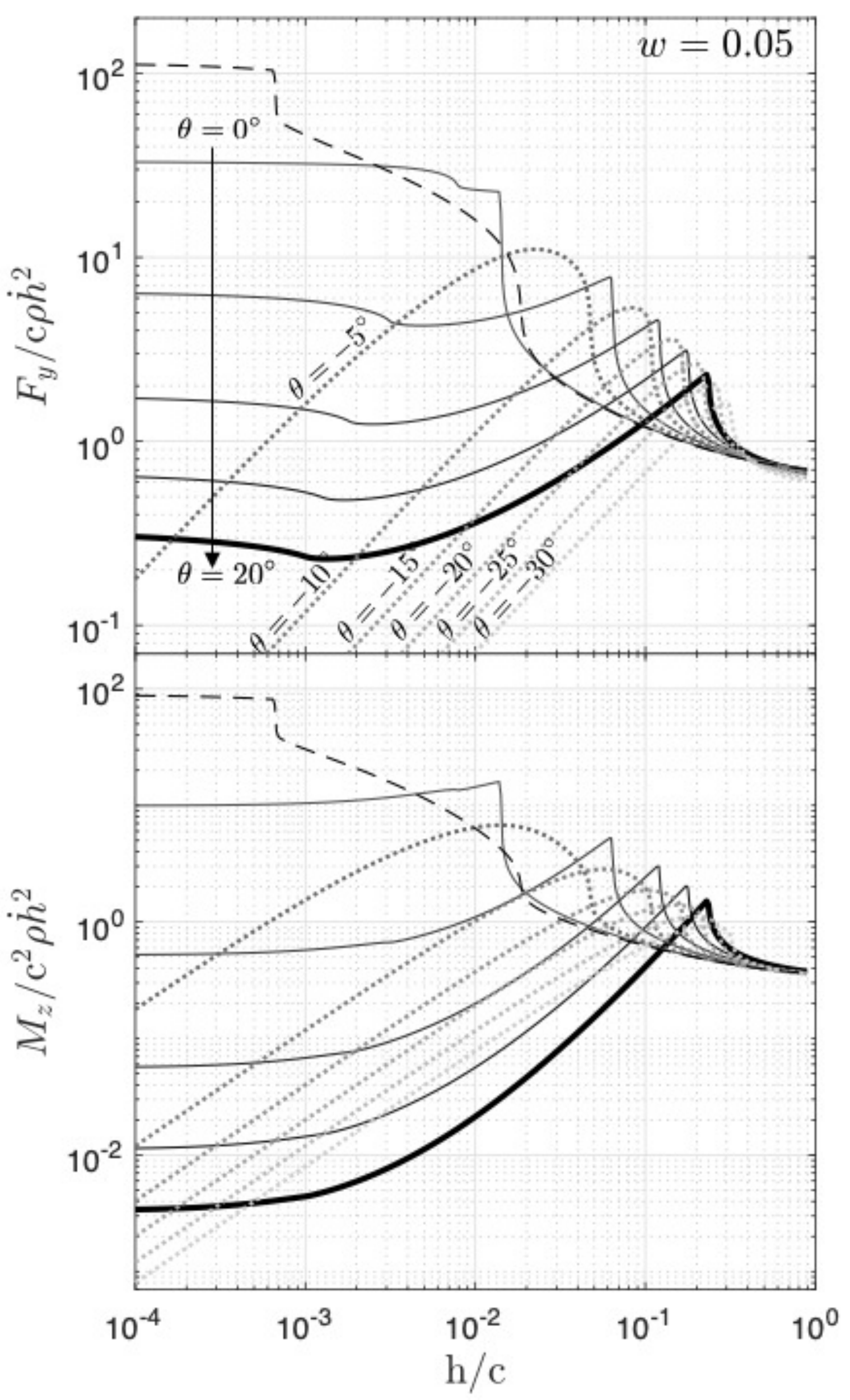}     
\end{tabular}
\end{center}
\caption{
Same as \fig{fig_naca_series_1} for different foil thicknesses: $w=0.1$ (left) and $w=0.05$ (right).
}
\label{fig_naca_series_2}
\end{figure}

% $$$$$$$$$$$$$$$$$$$$$$$$$$$$$$$$$$$$$$
% $$$$$$$$ FIGURES
% $$$$$$$$$$$$$$$$$$$$$$$$$$$$$$$$$$$$$$

\subsection{Slamming loads on foils of different thicknesses}
\label{subsec_diff_thick}

Using CFD results for comparison, the FBC concept has been shown to provide reliable estimates 
of the slamming loads on a NACA 0028 foil.
The present paragraph gives a more systematic description of the slamming loads acting on foils entering water, 
based on FBC estimates only.
Figs. \ref{fig_naca_series_1}-\ref{fig_naca_series_2} show the evolution of slamming loads 
as a function of the penetration depth $h$ and inclination angle $\theta$ for NACA foils of different thicknesses.
The inclination angle $\theta$ is restricted to $-30^{\circ} \rightarrow 20^{\circ}$,
range for which the FBC results have been compared to the CFD simulations in \S\ref{subsect_cfd_compare}.
 As the FBC estimates for $F_x$ have been found to be not as reliable as for $F_y$ and $M_z$, they are not shown in Figs. \ref{fig_naca_series_1}-\ref{fig_naca_series_2}.
This is not an issue for discussion purpose, since the magnitude of $F_x$ is much smaller than the magnitude of $F_y$.

\subsubsection{Initial slamming loads}

If the first contact occurs at a blunt point, the body contour can be locally approximated by a parabola $f(x) = x^2/2R_c$,
where $R_c$ is the local radius of curvature.
The half-width of the wetted region for a parabola is given by $\lambda = 2\sqrt{h R_c}$.
When $h\rightarrow0$ the expansion rate of the wetted correction becomes infinite: 
the MLM pressure reduces to linearised Bernoulli's relation (same argument as in \S\ref{subsec_fp_wagner}),
and the added-mass pressure becomes negligible.
This results in the following asymptotic value\footnote{
The asymptotic value is different for a body initially at rest on the free surface, suddenly accelerating into the fluid.
In this case, added-mass pressure is not negligible and one can show that $\lim\limits_{h \rightarrow 0} \frac{F_{y}(h)}{\rho \dot{h}^2} = 3\pi R_c$.
As an illustration, see \fig{fig_naca_acc} and compare initial values of $F_{y}/\rho \dot{h}^2$ on left and right panels.
} of $F_y$
\begin{equation}
\label{eq_fy_ini}
\lim\limits_{h \rightarrow 0} \frac{F_{y}(h)}{\rho \dot{h}^2} = \frac{F_{y}(0)}{\rho \dot{h}^2} = 2\pi R_c \, .
\end{equation}

In Figs. \ref{fig_naca_series_1}-\ref{fig_naca_series_2}, all 
the $F_y$ curves for $\theta>-\delta$ converge towards the asymptotic value given by \eq{eq_fy_ini}.
For $\theta<-\delta$, the first contact with the fluid occurs at the sharp trailing edge and $F_{y}(0) = 0$.
For a given foil thickness, the maximum values of $R_c$ and $F_{y}(0)/\rho \dot{h}^2$ 
are obtained for a negative inclination angle $\theta = \theta_m$ (values are given in Tab. \ref{tab_geometry}).
The slamming loads for $\theta = \theta_m$ are shown as dashed black lines in Figs. \ref{fig_naca_series_1}-\ref{fig_naca_series_2}.

% %%%%%%%%%%%%%%%%

\subsubsection{Time evolution of the slamming loads and flow separation}

For $\theta = 20^{\circ}$, flow separation first occurs at the leading edge and then at the trailing edge.
Starting from an initial value given by \eq{eq_fy_ini}, $F_y$ decreases and reaches a local minimum.
 Then, $F_y$ starts rising as the wetted area on the foil keeps increasing towards the trailing edge.
When the flow separates from the trailing edge, $F_y$ reaches a peak at $h/c \simeq 0.2-0.3$ and starts decreasing again. 
These peak values are local maxima for $w=0.3;0.2$, and global maxima for $w=0.1;0.05$.
Conversely, flow separation from the leading edge does not 
induce a sudden change of slope
in the evolution of $F_y$ and $M_z$,
as it occurs on a smooth part of the body.

Then as $\theta$ decreases from $20^{\circ}$ to $-30^{\circ}$,
separation at the leading edge and the trailing edge happen respectively later and earlier.
Separation happens simultaneously on both sides for a negative value of inclination angle, 
$\theta = \theta_s$ (values are given in Tab. \ref{tab_geometry}). 
For $-\delta < \theta  < \theta_s $, 
the separation from the trailing edge induces a sharp drop in $F_y$ and $M_z$,
whereas the separation from the leading edge still does not 
induce a sharp transition.
 
For $\theta < -\delta $, first contact with the fluid occurs at the trailing edge.
At early times, the wetted contour of the foil locally resembles an inclined flat plate; 
$F_y$ linearly increases with $h$.
As the fluid-body contact point approaches the leading edge, the deadrise angles increase resulting in a lower increase rate of $F_y$;
the vertical force $F_y$ starts decreasing slightly before the flow separation from the leading edge.

% %%%%%%%%%%%%%%%%

\section{Discussion}
\label{sect_conclusions}

We have shown that the Fictitious Body Continuation concept can be effectively used to 
estimate the hydrodynamic loads during the vertical water entry of two-dimensional asymmetric bodies,
including flow separation.
The FBC method is computationally fast and simple to implement compared to CFD or BEM methods.
Although the FBC model can properly mimic flow separation 
regarding hydrodynamic loads on the body,
we note that it does not provide a proper description of the free surface shape after flow separation.

In the present study, 
the real body contour is continued by fictitious flat plates.
Then the critical point in the model is the choice, \textit{a priori}, of the continuation angles $\alpha_1$ and $\alpha_2$.  
Comparisons with experiments or self-sufficient models (e.g. CFD simulations) 
are required to have some `heuristic' knowledge of suitable continuation angles.
This question was partly investigated by Tassin et al. (2014) \cite{tassin_2014}, considering simple symmetric body shapes.
They found as `best' continuation angles, 
$\alpha_{\rm fp} = 47^{\circ}$ for a horizontal flat-plate, $\alpha_{\rm cl} = 60^{\circ}$ for a circular cylinder, 
and $45^{\circ}-55^{\circ}$ for wedges with different deadrise angles.
For more general asymmetric bodies, one practical way of using the FBC concept
could be to find the appropriate `phenomenological' laws for continuation angles as a function of body orientation, 
$\alpha_1$ and $\alpha_2$. 
Such laws could be derived from the interpolation (or regression) of best-fit values obtained 
through comparisons with experiments or CFD simulations carried out for a few inclination angles. 

However, the present work shows that best continuation angles may weakly depend on the exact shape of the real body contour.
In Section \ref{sec_inclined_fp}, 
we have found that the continuation angle $\alpha_{\rm fp}$ is suitable to mimic the separated flow emerging from the tip of an inclined flat plate 
up to $\theta \simeq 30^{\circ}$. 
In this range of inclination angles, 
the FBC results agree with the nonlinear model of Faltinsen and Semenov \cite{faltinsen_semenov_2008} by $10\%$.
In Section \ref{sec_naca}, we have considered the vertical water entry of a foil, as a more complicated body shape.
We have set the continuation angles to $\alpha_1 = \alpha_{\rm cl}$ for flow separation at the smooth leading edge, and $\alpha_2 = \alpha_{\rm fp} $ at the sharp trailing edge.
Through comparisons 
with CFD simulations, we have shown that the FBC model provides good estimates of the slamming loads on the foil 
for a broad range of inclination angles ($-30^{\circ} < \theta < 20^{\circ}$); this, without any change in the values of $\alpha_1$ and $\alpha_2$.
From these encouraging results, one could wonder whether $\alpha_{\rm fp}$ and $\alpha_{\rm cl}$ --
for flow separation at a chine and from a smooth body part respectively -- can be used as generic continuation angles for a broad family of body shapes.
Comparative studies for other asymmetric bodies would be useful to better delimit the generic feature of continuation angles.

% %%%%%%%%%%%%%%%%%%%%%%%%%%%%%%%%%%%%%%%%%%
% %%%%%%%%%%%%%%%%%%%%%%%%%%%%%%%%%%%%%%%%%%
% 
% ACKNOWLEDGEMENTS
%
% %%%%%%%%%%%%%%%%%%%%%%%%%%%%%%%%%%%%%%%%%%
% %%%%%%%%%%%%%%%%%%%%%%%%%%%%%%%%%%%%%%%%%%

\section*{Acknowledgements}

This work was supported by the French National Agency for Research (ANR) 
and the French Government Defense procurement and technology agency  (DGA) 
[ANR-17-ASTR-0026 APPHY].

% %%%%%%%%%%%%%%%%%%%%%%%%%%%%%%%%%%%%%%%%%%
% %%%%%%%%%%%%%%%%%%%%%%%%%%%%%%%%%%%%%%%%%%
% 
% APPENDICES
%
% %%%%%%%%%%%%%%%%%%%%%%%%%%%%%%%%%%%%%%%%%%
% %%%%%%%%%%%%%%%%%%%%%%%%%%%%%%%%%%%%%%%%%%

% %%%%%%%%%%%%%%%%%%%%%%%%%%%%%%%%%%%%%%%%%%
% 
% APPENDIX A
%
% %%%%%%%%%%%%%%%%%%%%%%%%%%%%%%%%%%%%%%%%%%

\section*{Appendix A. No gravity assumption: limit of validity}

In Wagner's approach, the effect of gravity on the flow dynamics is neglected.
To assess the validity of this assumption, for a given configuration of water entry,
the acceleration of the liquid can be compared with the gravitational acceleration.
For the sake of simplicity, let us consider the case of a symmetric body ($\dot{\lambda}_1 = \dot{\lambda}_2 = \dot{\lambda}$), 
entering water at constant velocity.
Then, the acceleration field of the fluid, deriving from the velocity potential $\varphi^{(w)}$ (Eq. \ref{eq_plaque_plane}),
is given by
\begin{equation}
\label{eq_acc_field}
\varphi^{(w)}_{,xt}(x,0,t)  = -\dot{h} \frac{\dot{\lambda}}{\lambda} \frac{x/\lambda}{[1-(x/\lambda)^2]^{3/2}} \, . 
\end{equation}
In Wagner's approach, the deadrise angle is assumed to be small, $\beta \ll 1$, and the convective acceleration can be neglected to the leading order.
Besides, the only force considered in the fluid domain is the one due to the pressure gradient:
\eq{eq_acc_field} can also be obtained by considering the pressure force density $-\nabla P / \rho$,
from the linearised Bernoulli relation (Eq. \ref{eq_p_linear}).

In \eq{eq_acc_field}, the quantity
\begin{equation}
\dot{h}\frac{\dot{\lambda}}{\lambda} \sim \frac{\dot{h}^2}{h} = \frac{\dot{h}}{t}
\end{equation}
appears as the relevant scale for the fluid acceleration.
Then, gravity can be neglected if $\dot{h}/t \gg \beta g $, leading to the constraint 
\begin{equation}
\label{eq_tg}
t \ll t_g = \frac{\dot{h}}{\beta g} \, .
\end{equation}
The factor $1/\beta$ accounts for the fact that only the gravity component acting along streamlines (on the body surface) should be considered.
After flow separation from the body, the situation is different as the separated flow (especially jets) is ``free'' to plunge and can impact on the underlaying liquid surface.
Besides, the cavity flow can begin to collapse; see Bao et al. \cite{bao_2016} for BEM simulations of the water entry of a finite wedge at different velocities.
Then, it may be safer to consider \eq{eq_tg} without the factor $1/\beta$.

To our knowledge, few studies did a systematic study of gravity effect in the water entry problem.
The pioneering work of Mackie \cite{mackie_1965} was maybe the first to focus on the effect of gravity in the water entry problem.
However, his analytical developments make the assumption of very large deadrise angles ($\beta$ close to $90^{\circ}$);
i.e. the extreme opposite of Wagner's assumption.
Zekri \cite{zekri_2016} did a perturbation study of gravity effect within the Wagner framework, 
for a body of parabolic shape, $f(x) = x^2/2R$.
His analysis yields the characteristic timescale $t_g = (R\dot{h}/g^2)^{1/3}$, which is identical to \eq{eq_tg} 
when taking $\beta = \sqrt{\dot{h} t_g/R}$ as the characteristic deadrise angle of the wetted parabolic contour at $t=t_g$.
However, his perturbative expansion is restricted to the early stage of water impact ($t/t_g \ll 1$), 
when the effect of gravity can still be neglected for practical concern.
Yan and Liu \cite{yan_2011} used a boundary element method to investigate the effect of gravity during the water impact of inverted cones,
with different deadrise angles. 
They did find that $\dot{h} / \beta g t$ could be used as a similarity parameter to describe the effect of gravity across their simulations.
Defining as a Froude number, $Fr = \sqrt{\dot{h}/gt}$, 
their simulations predict that gravity induces a $10\%$ increase in the impact load (hydrodynamic + hydrostatic)
at $Fr \simeq 0.44;0.98;1.7;2.6$ for $\beta = 15^{\circ};30^{\circ};45^{\circ};60^{\circ}$.

\section*{Appendix B. Computation of the Wagner wetted area}

This section gives some details about the scheme 
implemented to solve the system of equations \ref{wagner_condi_2d_left}-\ref{wagner_condi_2d_right}.
The function whose root is searched for is the following:
\begin{equation}
\label{eq_F_newton}
\begin{array}{c|c}
G(b_1, b_2) =  & 
\displaystyle g_1(b_1, b_2) =  \int_{-b_1}^{b2} f(x) \sqrt{\frac{b_2-x}{b_1+x}} \ {\rm d} x  -  \frac{\pi}{2} (b_1 + b_2) h  \\
& \\
 & 
\displaystyle g_2(b_1, b_2) = \int_{-b_1}^{b2} f(x) \sqrt{\frac{b_1+x}{b_2-x}} \ {\rm d} x -  \frac{\pi}{2} (b_1 + b_2) h  \, ,
\end{array}
\end{equation}
where $h$ is a parameter.
The root, $\Lambda = (\lambda_1$, $\lambda_2)$, is searched by using the Newton-Raphson method with a relaxation condition.
Starting from an initial guess $\Lambda^{0}$, the approximate value is iteratively improved by using the formula:
\begin{equation}
\Lambda^{i+1} = \Lambda^{i} - \omega \left[J_G\left(\Lambda^{i}\right)\right]^{-1} G\left(\Lambda^{i}\right) \,
\end{equation}
where $\omega$ is a relaxation factor; $\omega = 1$ corresponds to no relaxation. 
$J_G$ is the Jacobian matrix of G, given by
\begin{equation}
J_G = \left[
\begin{array}{cc}
\displaystyle \frac{\partial g_1}{\partial b_1} &  \displaystyle \frac{\partial g_1}{\partial b_2} \\
\vspace{-0.4cm}
&\\
\displaystyle \frac{\partial g_2}{\partial b_1} &  \displaystyle \frac{\partial g_2}{\partial b_2}
\end{array} 
\right] \, .
\end{equation}
For some configurations, the Newton-Raphson scheme did not converge without under-relaxation.
An under-relaxation factor $\omega = 0.1$ was found to provide convergence for all considered cases.

\noindent 
\textbf{Initial guess $\Lambda^{0}$.} 
The von Karman wetted area is used as an initial guess for $\lambda_1$ and $\lambda_2$.
The von Karman solution is found by looking for the intersection points between the body contour and the initial free surface. \\

\noindent
\textbf{Stopping criterion.}
The iterative process is stopped when the scalar quantity
\begin{equation}
r(\lambda_1^{i}, \lambda_2^{i}) = \frac{ \sqrt{g_1\left(\lambda_1^{i}, \lambda_2^{i}\right)^2+g_2\left(\lambda_1^{i}, \lambda_2^{i}\right)^2} }{\left(\lambda_1^{i}+\lambda_2^{i}\right)h}
\end{equation}
drops below a given threshold $\varepsilon$. For the results reported in the present study,
the threshold was set to $\varepsilon = 10^{-4}$.\\

\noindent \textbf{Computation of $G$.}
The body contour is approximated by a polygon.
In order to reduce the discretization error for a given number of discretization points, 
the vertices of the polygon are uniformly distributed as a function of the variable
\begin{equation}
\Gamma(s) = \int |\gamma(s)| d s \, ,
\end{equation}
where $s$ is the curvilinear abscissa along the body contour, and $\gamma$ its curvature.
The function representing the polygon, $\widehat{f}(x)$, is piecewise-linear.
Making the variable substitutions
\begin{equation}
\label{eq_xi}
\xi_{\pm} = \pm[2x-(b_2-b_1)]/(b_1+b_2) \, ,
\end{equation} 
the integrals appearing in \eq{eq_F_newton} become
\begin{equation}
\label{eq_I2}
\begin{array}{l}
\displaystyle I_-(b_1, b_2) = \int_{-b_1}^{b2} \widehat{f}(x) 
\sqrt{\frac{b_2-x}{b_1+x}} 
\ {\rm d} x   
= \frac{1}{2}(b_1+b_2)  \int_{-1}^{1} \widehat{f}[x(\xi_{-})] \sqrt{\frac{1+\xi_{-}}{1-\xi_{-}}} \ {\rm d} \xi_{-} \, .
\end{array}
\end{equation} 
\begin{equation}
\label{eq_I1}
\begin{array}{l}
\displaystyle I_+(b_1, b_2) =  \int_{-b_1}^{b2} \widehat{f}(x) 
\sqrt{\frac{b_1+x}{b_2-x}} 
\ {\rm d} x
= \frac{1}{2}(b_1+b_2) \int_{-1}^{1} \widehat{f}[x(\xi_{+})] \sqrt{\frac{1+\xi_{+}}{1-\xi_{+}}} \ {\rm d} \xi_{+}  \\
\end{array}
\end{equation} 
\eq{eq_xi} being a linear relationship between $\xi_\pm$ and $x$, $\widehat{f}[x(\xi_{+})]$ and $\widehat{f}[x(\xi_{-})]$ are also piecewise-linear functions.
Then, by using the antiderivatives
\begin{equation}
\label{eq_II0}
\begin{array}{l}
\displaystyle \int  \sqrt{\frac{1+u}{1-u}} \ {\rm d} u = -\sqrt{1-u^2} + 2 \arcsin{\sqrt{(1+u)/2}}
\end{array}
\end{equation} 
\begin{equation}
\label{eq_II1}
\begin{array}{l}
\displaystyle \int  u \sqrt{\frac{1+u}{1-u}} \ {\rm d} u = -(u/2+1)\sqrt{1-u^2} + \arcsin{\sqrt{(1+u)/2}} \, ,
\end{array}
\end{equation} 
the integrals appearing in Eqs. (\ref{eq_I2}-\ref{eq_I1}) can be expressed as the sum of analytical terms,
which are numerically added up.
The partial derivatives required to compute the Jacobian matrix, $J_G$, 
are numerically obtained by using a perturbation method.

\bibliography{mybibfile}

\end{document}